\documentclass[%
reprint,
superscriptaddress,
aps,
pra,
]{revtex4-2}

\usepackage{graphicx}% Include figure files
\usepackage{dcolumn}% Align table columns on decimal point
\usepackage{bm}% bold math
\usepackage{physics}
\usepackage[caption=false]{subfig}
\usepackage{hyperref}% add hypertext capabilities
\usepackage{color}
\usepackage{ulem}
\usepackage{float}

\begin{document}

%\preprint{APS/123-QED}

\title{Spectral functions and localization landscape theory in speckle potentials}

\author{Pierre Pelletier}
\affiliation{
Laboratoire de Physique de la Matière Condensée, Ecole Polytechnique, CNRS, Institut Polytechnique de Paris, 91120 Palaiseau, France
}

\author{Dominique Delande}
\affiliation{
Laboratoire Kastler Brossel, Sorbonne Université, CNRS, ENS-PSL Research University,
Collège de France, 4 Place Jussieu, 75005 Paris, France
}

\author{Vincent Josse}
\affiliation{
Laboratoire Charles Fabry, Institut d’Optique, CNRS, Université Paris-Saclay, 91127 Palaiseau cedex, France
}

\author{Alain Aspect}
\affiliation{
Laboratoire Charles Fabry, Institut d’Optique, CNRS, Université Paris-Saclay, 91127 Palaiseau cedex, France
}

\author{Svitlana Mayboroda}
\affiliation{
School of Mathematics, University of Minnesota, Minneapolis, Minnesota 55455, USA
}

\author{Douglas Arnold}
\affiliation{
School of Mathematics, University of Minnesota, Minneapolis, Minnesota 55455, USA
}

\author{Marcel Filoche}
\affiliation{
Laboratoire de Physique de la Matière Condensée, Ecole Polytechnique, CNRS, Institut Polytechnique de Paris, 91120 Palaiseau, France
}

\date{\today}

\begin{abstract}
Spectral function is a key tool for understanding the behavior of Bose-Einstein condensates of cold atoms in random potentials generated by a laser speckle. In this paper we introduce a new method for computing the spectral functions in disordered potentials. Using a combination of the Wigner-Weyl approach with the landscape theory, we build an approximation for the Wigner distributions of the eigenstates in the phase space and show its accuracy in all regimes, from the deep quantum regime to the intermediate and semiclassical. Based on this approximation, we devise a method to compute the spectral functions using only the landscape-based effective potential. The paper demonstrates the efficiency of the proposed approach for disordered potentials with various statistical properties without requiring any adjustable parameters.
\end{abstract}

\maketitle

%\tableofcontents

\section{Introduction}

Due the very weak interactions with the environment and the tunability of internal interactions, cold atom systems are of particular interest to study the influence of disorder on quantum transport~\cite{Aspect2009, Shapiro2012}. Cold atoms in the disordered potential of a laser speckle provided the first direct observation of Anderson localization~\cite{Billy2008}, theoretically predicted in 1958~\citep{Anderson1958}. In these systems, the so-called \emph{spectral functions} $A_\mathbf{k}(E)$ are quantities of special importance both for theory and experiments. They are defined as the (averaged over disorder realizations) energy distribution of a quantum plane wave of wave vector $\mathbf{k}$ (hence of a quantum particle of momentum~$\hbar \mathbf{k}$). Alternatively, $A_\mathbf{k}(E)$ can be seen as the average momentum distribution of a state with energy $E$. The spectral function is a building block for computing properties of disordered quantum systems. For example, the density of states is simply the integral of the spectral function over $\mathbf{k}$. In three-dimensional (3D) systems, there is generically a transition between Anderson localized states at low energy and delocalized states at high energy. This second order continuous phase transition---known as the Anderson transition---takes place at a given energy $E_c$ called the mobility edge. Computing $E_c$ and the (universal) properties in its vicinity for a given system is a difficult task, mainly because it takes place in the strong disorder regime where perturbative approaches fail. One could expect the spectral function to present some singularity at $E_c,$ but this turns out to be too naive. The main reason is that the spectral function involves---see below---a single averaged Green function while the transport properties which are singular at~$E_c$ involve the average product of a retarded and an advanced Green functions~\cite{Kuhn2007, Piraud2013, AkkermansMontambaux:MesoscopicPhysics:11}.

Non-perturbative approaches have been tried to predict the position of the mobility edge, one of the more advanced ones being the self-consistent theory of localization~\cite{Vollhardt1992}. It has been used for the specific case of cold atoms in a disordered optical potential in \cite{Kuhn2007, Yedjour2010, Piraud2013}. However, the results depend on the approximations, and only a semi-quantitative agreement between the experimental results~\cite{Jendrzejewski2012, Semeghini2015} and the numerical calculations~\cite{Delande2014, Pasek2017} is obtained for the position of the mobility edge. In any case, a crucial ingredient for estimating the mobility edge in self-consistent theories is the spectral function. This is because the transport properties are measured in the momentum space while the mobility edge is in the energy space; the spectral function provides the necessary link between spatial and energetic properties. This is why a number of studies have been devoted to compute these spectral functions~\cite{Trappe2015, Prat2016}, as well as to measure them experimentally~\cite{Volchkov2018}.

In this paper, we present a new approach to computing the spectral functions based on a recent theoretical development on localization called the \emph{Localization Landscape} (LL) theory. After a short presentation of the main properties of the localization landscape and of its reciprocal, the effective potential, we introduce the computation of the spectral functions based on the Weyl transform of a Hamiltonian derived from the LL and on the Wigner functions of quantum eigenstates. This allows us to showcase how the LL provides a much more accurate view of the distribution of these Wigner functions than the original Hamiltonian. Finally we compute spectral functions using this new approach for a variety of potentials including laser speckle, and assess the accuracy of our approach both in the semiclassical and quantum regimes.

\section{The Localization Landscape theory}

The LL is a theoretical tool that was shown to possess information regarding localization of waves in a large variety of physical systems~\cite{Filoche2012, Arnold2016, Filoche2017, Arnold2019a, Arnold2019b, David2021}. We will briefly introduce its definition and properties in the context of a single quantum particle in a non-negative potential $V$. In fact, the potential only needs to be bounded from below, in which case it can be uplifted by a constant to become positive. The Hamiltonian is therefore $\hat{H} = -\frac{\hbar^2}{2m} \Delta + \hat{V}$, and the LL~$u$ is defined as the solution to
\begin{equation}\label{eq:landscape}
    \hat{H}u \,= -\frac{\hbar^2}{2m} \Delta u+ V u = 1 \,,
\end{equation}
with the boundary conditions corresponding to those of the original problem. The function~$u$ is remarkably simple to compute numerically considering the large amount of information it provides. First, to address those, recall that an essential role is played by the reciprocal of the LL, $V_u \equiv 1/u$ \footnote{Please note that, contrary to previous papers, this potential had been denoted $V_u$ instead of $W$ to avoid any confusion with the Wigner function}, which acts as an \emph{effective potential} through the following identity satisfied by any state~$\ket{\psi}$~\cite{Arnold2016}:
\begin{equation}
    \ev{\hat{H}}{\psi} =  \frac{\hbar^2}{2m} \bra{u\nabla\left(\frac{\psi}{u}\right)}\ket{u\nabla\left(\frac{\psi}{u}\right)} + \bra{\psi}\hat{V}_u\ket{\psi} \,.
\end{equation}
This identity shows that the energy of any state can be decomposed into two positive contributions, the first one that is akin to an effective kinetic energy, and the second one which is the potential energy of the state in the potential~$V_u$. One of the key properties of $V_u$ is that it can be used to build a very good approximation to the integrated density of states (the number of states below a given energy, denoted here by IDOS)~\cite{Arnold2016, Arnold2019b, David2021}. One of these approximations is inspired by Weyl's asymptotic law. The classical Weyl law provides a high-energy asymptotic expression of the IDOS by counting the volume in phase space accessible for a particle of energy $E$, i.e., the volume enclosed within the hypersurface of equation $H(\vb{x},\vb{p})=E$, where $H(\vb{x},\vb{p})$ is the classical Hamiltonian function. In the following, for convenience, we will use $\vb{k}$ instead of the momentum $\vb{p}=\hbar \vb{k}$. The LL-based approximation uses a similar expression, only replacing the original potential by the effective potential~$V_u$:
\begin{align}\label{eq:N_Vu}
    & IDOS(E) \approx \frac{1}{(2\pi)^d} \iint_{H_1(\vb{x},\vb{k})\leq E}  \mathrm{d}\vb{x}\, \mathrm{d}\vb{k} \\
    &\textrm{with} \qquad H_1(\vb{x},\vb{k}) = \frac{\hbar^2 \vb{k}^2}{2m} + V_u(\vb{x})\,.
\end{align}
The accuracy of this approximation even at low energy, demonstrated in ~\cite{Arnold2016, Arnold2019b}, suggests that the eigenfunctions of $\hat{H}$ of energy comprised in the interval $[E, E+dE]$ are to be ``found'' in phase space mostly between the hypersurfaces of equations $H_1(\vb{x},\vb{k})=E$ and $H_1(\vb{x},\vb{k})=E+dE$, respectively. In order to make our statement more precise, we now introduce the Wigner-Weyl formalism establishing a link between the Hilbert space of quantum states and Hermitian operators on one side, and distributions and functions in phase space on the other side.

\section{The Wigner-Weyl approach} 

\subsection{The Wigner-Weyl formalism} 

The\emph{ Wigner function} of a wave function~$\psi$ is defined in the phase space~$(\vb{x},\vb{k})$ as \cite{Wigner1932,Hillery1984}
\begin{equation}
	\label{eq:Wigner}
    W_\psi(\vb{x},\vb{k}) \equiv \frac{1}{(2\pi)^d} \int e^{-i \vb{k} \cdot \vb{y}} \,\psi^*\left(\vb{x}-\frac{\vb{y}}{2}\right) \, \psi\left(\vb{x}+\frac{\vb{y}}{2}\right)\,\mathrm{d}\vb{y} 
\end{equation}
It can also be defined as:
\begin{equation}
    W_\psi(\vb{x},\vb{k}) \equiv \frac{1}{(2\pi)^d} \int e^{i \vb{k}' \cdot \vb{x}} \,\chi^*\left(\vb{k}+\frac{\vb{k}'}{2}\right) \, \chi\left(\vb{k}-\frac{\vb{k}'}{2}\right)\,\mathrm{d}\vb{k}'
\end{equation}
where $\chi$ is the Fourier transform of the wave function defined as
\begin{equation}
    \chi(\vb{k}) = \frac{1}{(2\pi)^{d/2}} \int e^{-i \vb{k} \cdot \vb{x}} \,\psi(\vb{x}) \, \mathrm{d}\vb{x}  \,.
\end{equation}

The Wigner function satisfies many important properties. It is dimensionless and can be understood, in particular, as being close to a probability distribution in phase space~\cite{Case2008}. This last statement comes from the fact that the two marginal integrals along~$\vb{x}$ and~$\vb{k}$ satisfy
\begin{align}
    \int W_\psi(\vb{x},\vb{k}) \, \mathrm{d}\vb{k} &= \abs{\psi(\vb{x})}^2 \,, \label{eq:margin_x}\\
    \int W_\psi(\vb{x},\vb{k}) \, \mathrm{d}\vb{x} &= \abs{\chi(\vb{k}) }^2\,.\label{eq:margin_k}
\end{align}

In addition, the Hermitian inner product in Hilbert space is transported into the inner product on distributions in phase space through the following identity satisfied by any pair of quantum states $(\psi_1$, $\psi_2)$:
\begin{equation}
%    \int_{(\vb{x},\vb{k})} W_{\psi_1}(\vb{x},\vb{k}) \, W_{\psi_2}(\vb{x},\vb{k}) \, \mathrm{d}\vb{x} \, \mathrm{d}\vb{k} = \abs{\braket{\psi_1}{\psi_2}}^2 \,.
    \iint W_{\psi_1}(\vb{x},\vb{k}) \, W_{\psi_2}(\vb{x},\vb{k}) \, \mathrm{d}\vb{x} \, \mathrm{d}\vb{k} = \abs{\braket{\psi_1}{\psi_2}}^2 \,.
\end{equation}
One should note however that $W(\vb{x}, \vb{k})$ is not a genuine probability distribution over phase-space~\cite{Hillery1984}. For instance, it can take negative values (although it is positive when convolved with any with any minimal Gaussian such that $\Delta x\Delta k=1/2$). Consequently, one should be careful in the physical interpretation of this quantity.

This formalism is completed on the operator side by the \emph{Weyl transform}~\cite{Weyl1931}. The Weyl transform of any operator $\hat{O}$ acting on quantum states is defined as
\begin{equation}
    \widetilde{O}(\vb{x},\vb{k}) = \int e^{-i \vb{k} \cdot \vb{y}} \mel{\vb{x}+\frac{\vb{y}}{2}}{\hat{O}}{\vb{x} - \frac{\vb{y}}{2}} \,\mathrm{d}\vb{y} \,.
\end{equation}
For instance, the Weyl transform of the Hamiltonian~$\hat{H}$ is $\widetilde{H}(\vb{x},\vb{k}) = \hbar^2 \vb{k}^2/2m + V(x)$ [similarly, $\widetilde{H}_1(\vb{x},\vb{k}) = \hbar^2 \vb{k}^2/2m + V_u(\vb{x})$]. For sake of simplicity, we will omit the tilde symbol on $H(\vb{x},\vb{k})$ and $H_1(\vb{x},\vb{k})$ when referring to the Weyl transforms of these operators in the rest of the paper.

The main property of the Weyl transform, in conjunction with the Wigner function, is that it provides a measure on phase space which can be used to compute the expectation of any observable through the following identity:
\begin{equation}\label{eq:expval}
    \expval{O}_\psi = \ev{\hat{O}}{\psi} = \iint W_\psi(\vb{x},\vb{k}) \,\widetilde{O}(\vb{x},\vb{k}) \,\mathrm{d}\vb{x} \, \mathrm{d}\vb{k}\,.
\end{equation}
In short, the Wigner-Weyl formalism offers a way to envision any observable quantity in phase space. We will now see how the approximation~\eqref{eq:N_Vu} reveals the structure of quantum states in phase space, especially in the case of disordered potentials.

\subsection{The structure of quantum states in phase space} 

By definition, the IDOS counts the number of quantum states below energy~$E$. The accuracy of approximation~\eqref{eq:N_Vu} (cf. \citep{Arnold2016}) suggests that, in phase space, the supports of the Wigner functions of the quantum states in the energy range $[E, E+dE]$ are located mostly between the two hypersurfaces $H_1(\vb{x},\vb{k})=E$ and $H_1(\vb{x},\vb{k}) = E+dE$.

To test this hypothesis, we now consider a Hamiltonian with a statistically translation-invariant disordered potential $V$. We denote by $\overline{V}$ its average value, by $P(V)$ its probability density, and by $g(\vb{x})$ its spatial correlation function defined as
\begin{equation}
    g(\vb{x}) = \overline{V(\vb{x'})V(\vb{x'}+\vb{x}) } - \overline{V(\vb{x'})}^2\,,
\end{equation}
where $\overline{\cdots}$ corresponds to averaging over the statistical ensemble of disordered potentials. In the rest of the paper, we will use non bold notations $x$ and $k$ instead of $\vb{x}$ and $\vb{k}$ when dealing with one-dimensional systems since these quantities are scalar in this case. We use an ensemble especially relevant for cold atoms, the \textit{blue-detuned speckle potential} with a correlation function approximated by a Gaussian. This random potential is characterized by:
\begin{align} \label{eq:speckleGaussian}
\begin{cases}
    P(V) &= \displaystyle \frac{1}{V_0} \exp(-\frac{V}{V_0}) \,\theta(V) \,,\\
    \overline{V} &= V_0 \,,\\
    g(x) &= {V_0}^2 ~ \exp\left(-\displaystyle \frac{x^2}{2\sigma^2}\right) \,,
\end{cases}
\end{align}
where $\theta$ is the Heavyside step function and $\sigma$ is the correlation length of the potential. This length~$\sigma$ defines another important energy scale, the correlation energy~\cite{Kuhn2007}
\begin{equation}
    E_\sigma = \frac{\hbar^2}{m\sigma^2} \,,
\end{equation}
which corresponds to the typical energy of a particle with a de Broglie wavelength of the order of~$\sigma$. We can then define the ratio between the two energy scales $E_\sigma$ and $V_0$
\begin{equation}
    \eta = \frac{V_0}{E_\sigma} = \frac{m\sigma^2 V_0}{\hbar^2} \,.
    \label{eq:eta}
\end{equation}
Depending on the value of this dimensionless parameter, the system can explore a semiclassical regime ($\eta \gg 1$), a deeply quantum regime ($\eta \ll 1$), or an intermediate regime ($\eta \approx 1$)~\cite{Falco2010}. This parameter plays a key role in understanding the behavior of the system. Moreover, when exploring the values of~$\eta$ from the semiclassical to the quantum regime, the effective potential~$V_u$ is expected to  gradually change from the original potential~$V$ to a renormalized disorder that accounts for quantum confinement and tunneling effects.

In the rest of the paper, we use $\sigma$ as the unit of length, and set $m = \hbar = 1$ such that we have simply $\eta = V_0$ and $H(x,k) = k^2/2 + V(x)$.

In the following plots, we analyze the ``essential'' supports of the Wigner functions of the eigenfunctions of the Hamiltonian in phase space, for different values of the disorder strength~$V_0$ (or parameter~$\eta$) and at various energies. For each target energy $E$, all consecutive eigenstates in the interval $[E-\Delta E, E+\Delta E]$ are considered, with $\Delta E = \alpha E$ and $\alpha=0.2$. The eigenfunctions are computed using a finite difference scheme by discretizing the Hamiltonian~$\hat{H}$ on a grid of step~$\delta$ so that the discretized Schrödinger equation at energy~$E$ reads
\begin{equation}
    \frac{2 \psi^{(n)} - \psi^{(n-1)} - \psi^{(n+1)}}{2 \delta^2} + V_n \psi^{(n)} = E \, \psi^{(n)}
\end{equation}
where $\psi^{(n)} = \psi(n\delta)$ and $V_n = V(n\delta)$. In order for the solution $(\psi^{(n)})$ to be a good approximation of the continuous one, $\delta$ must be chosen significantly smaller than both the correlation length $\sigma$ and the de Broglie wavelength $\Lambda \propto 1/\sqrt{E}$. In our simulations, the system length is $L=200$ and $\delta =0.2$, and the boundary conditions are periodic. In summary, for each realization of the speckle (disordered) potential, we compute the associated LL function, all~eigenstates $(\psi_i)$ around a target energy~$E$ in the interval $[E-\Delta E, E + \Delta E]$, as well as the corresponding Wigner functions~$W_i(x,k) = W_{\psi_i}(x,k)$~
\footnote{In a discretized configuration space with periodic boundary conditions, the integral defining the Wigner function, Eq.~\eqref{eq:Wigner}, reduces to a discrete sum which is easily computed using a fast Fourier transform~\cite{Kolovsky:Wigner:Chaos1996, Saraceno:Wigner:PRA2002, Arguelles2005}. A slight complication is the appearance of ``ghost images''~\cite{Arguelles2005} of the interesting structures, which we eliminated using the method proposed in Ref.~\cite{Kolovsky:Wigner:Chaos1996}.}.

\begin{figure}[ht!]
\includegraphics[scale = 0.5]{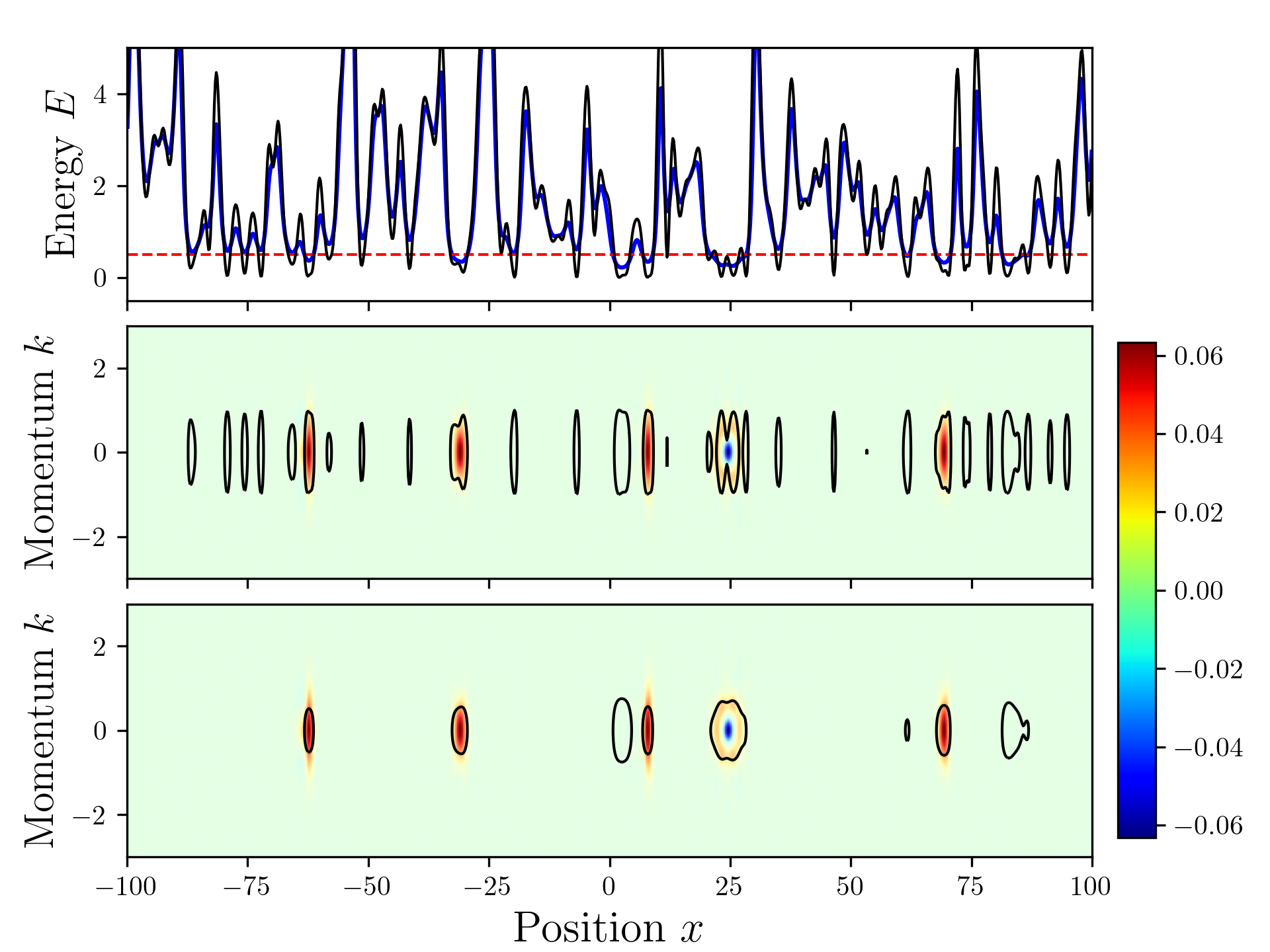}
\caption{\textbf{Semiclassical regime}. Top frame: disordered potential~$V$ with $\eta=2$ (black line), effective potential $V_u$ (blue line) and energy level at $E=0.5$ (dashed red line). Middle frame: color plot of $F_E(x,k)$ defined in Eq.~\eqref{eq:FE} computed over all eigenfunctions in a range $\Delta E = 0.2 E$ around the target energy $E=0.5$, superimposed with the level set $H(x,k)=E$ represented by a continuous black line. Bottom frame: Similar representation as in the middle frame, but with the level set $H_1(x,k)=E$.}
\label{fig:eta2}
\end{figure}

\begin{figure}[ht!]
\includegraphics[scale = 0.5]{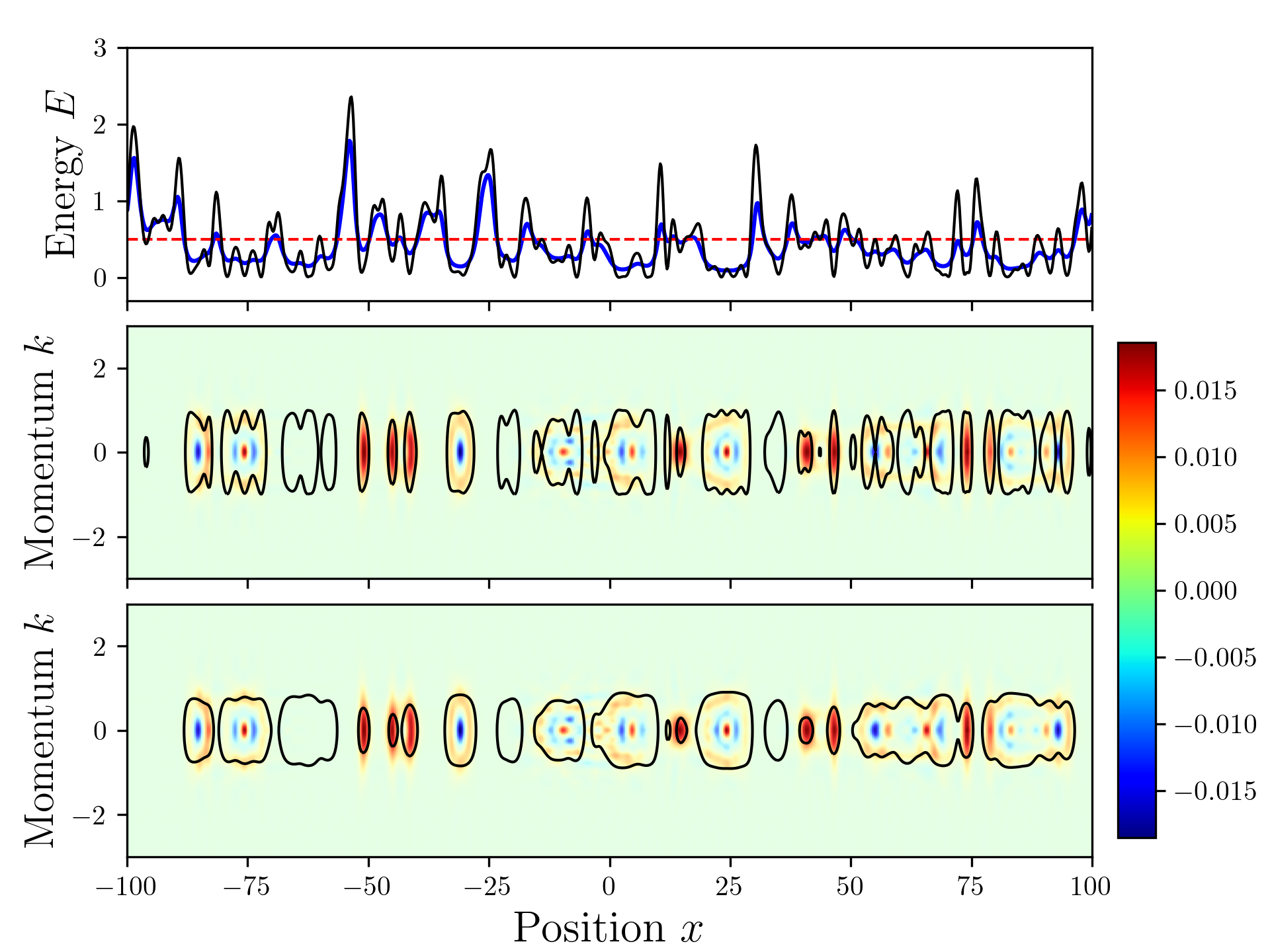}
\caption{\textbf{Quantum regime}. Top frame: disordered potential~$V$ with $\eta=0.5$ (black line), effective potential $V_u$ (blue line) and energy level at $E=0.5$ (dashed red line). Middle frame: color plot of $F_E(x,k)$ defined in Eq.~\eqref{eq:FE} computed over all eigenfunctions in a range $\Delta E = 0.2 E$ around the target energy $E=0.5$, superimposed with the level set $H(x,k)=E$ represented by a continuous black line. Bottom frame: Similar representation as in the middle frame, but with the level set $H_1(x,k)=E$.}
\label{fig:eta05}
\end{figure}

\begin{figure}[ht!]
\includegraphics[scale = 0.5]{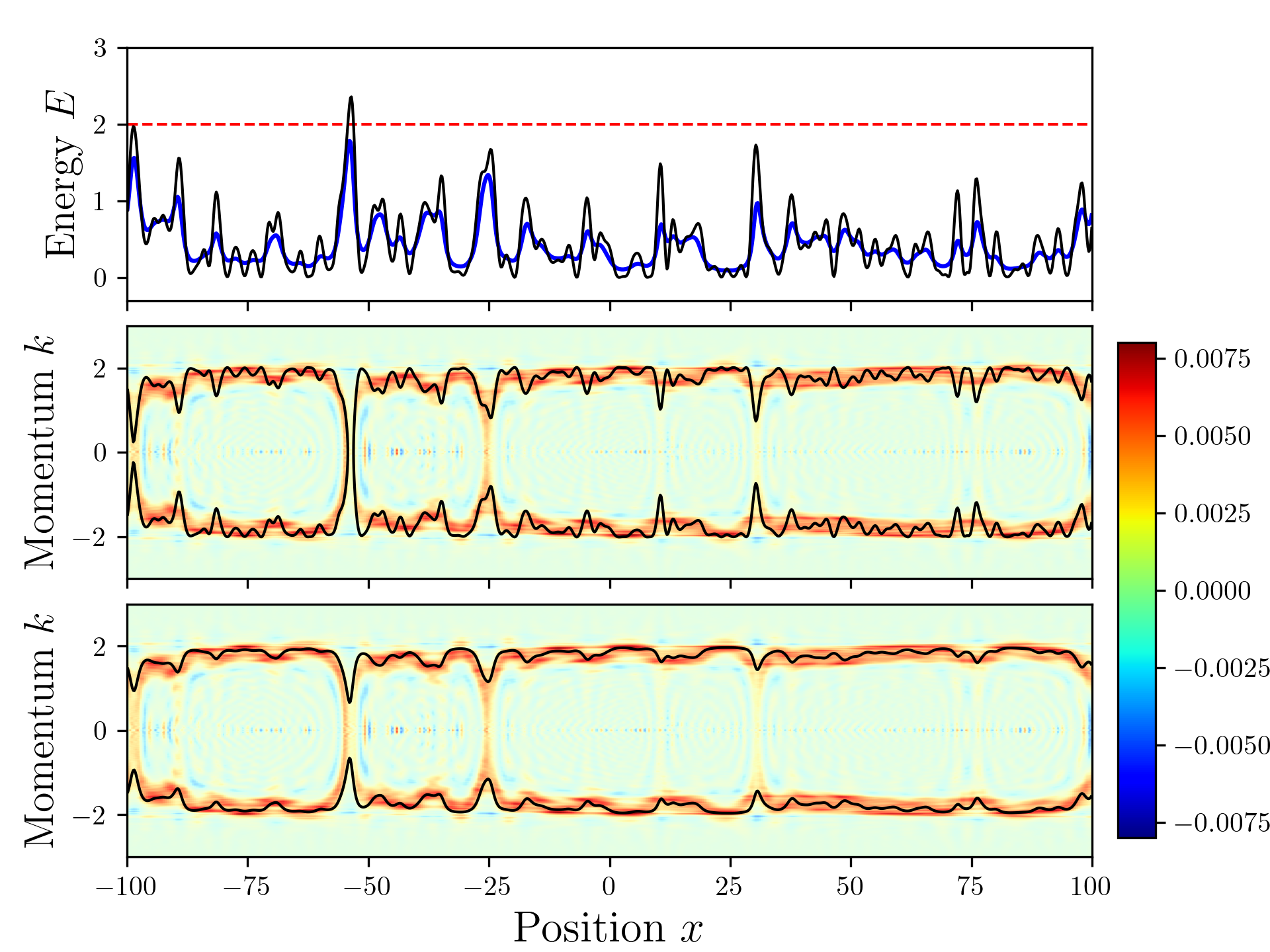}
\caption{\textbf{Quantum regime}. Top frame: disordered potential~$V$ with $\eta=0.5$ (black line), effective potential $V_u$ (blue line) and energy level at $E=2$ (dashed red line). Middle frame: color plot of $F_E(x,k)$ defined in Eq.~\eqref{eq:FE} computed over all eigenfunctions in a range $\Delta E = 0.2 E$ around the target energy $E=2$, superimposed with the level set $H(x,k)=E$ represented by a continuous black line. Bottom frame: Similar representation as in the middle frame, but with the level set $H_1(x,k)=E$.}\label{fig:eta05E2}
\end{figure}

\begin{figure}[ht!]
\includegraphics[scale = 0.5]{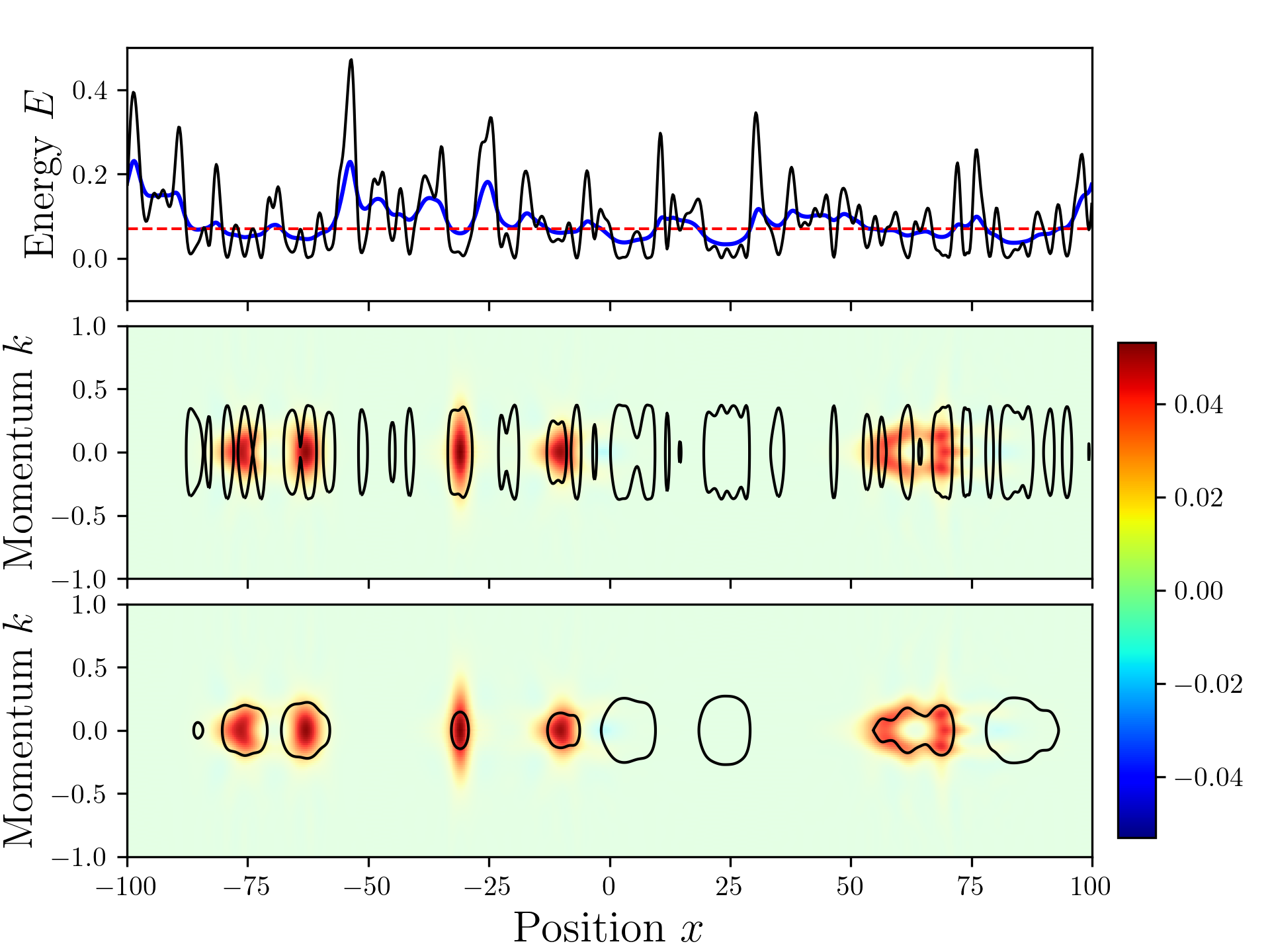}
\caption{\textbf{Deep quantum regime}. Top frame: disordered potential~$V$ with $\eta=0.1$ (black line), effective potential $V_u$ (blue line) and energy level at $E=0.07$ (dashed red line). Middle frame: color plot of $F_E(x,k)$ defined in Eq.~\eqref{eq:FE} computed over all eigenfunctions in a range $\Delta E = 0.2 E$ around the target energy $E=0.07$, superimposed with the level set $H(x,k)=E$ represented by a continuous black line. Bottom frame: Similar representation as in the middle frame, but with the level set $H_1(x,k)=E$.}
\label{fig:eta01lowest}
\end{figure}

\begin{figure}[ht!]
\includegraphics[scale = 0.5]{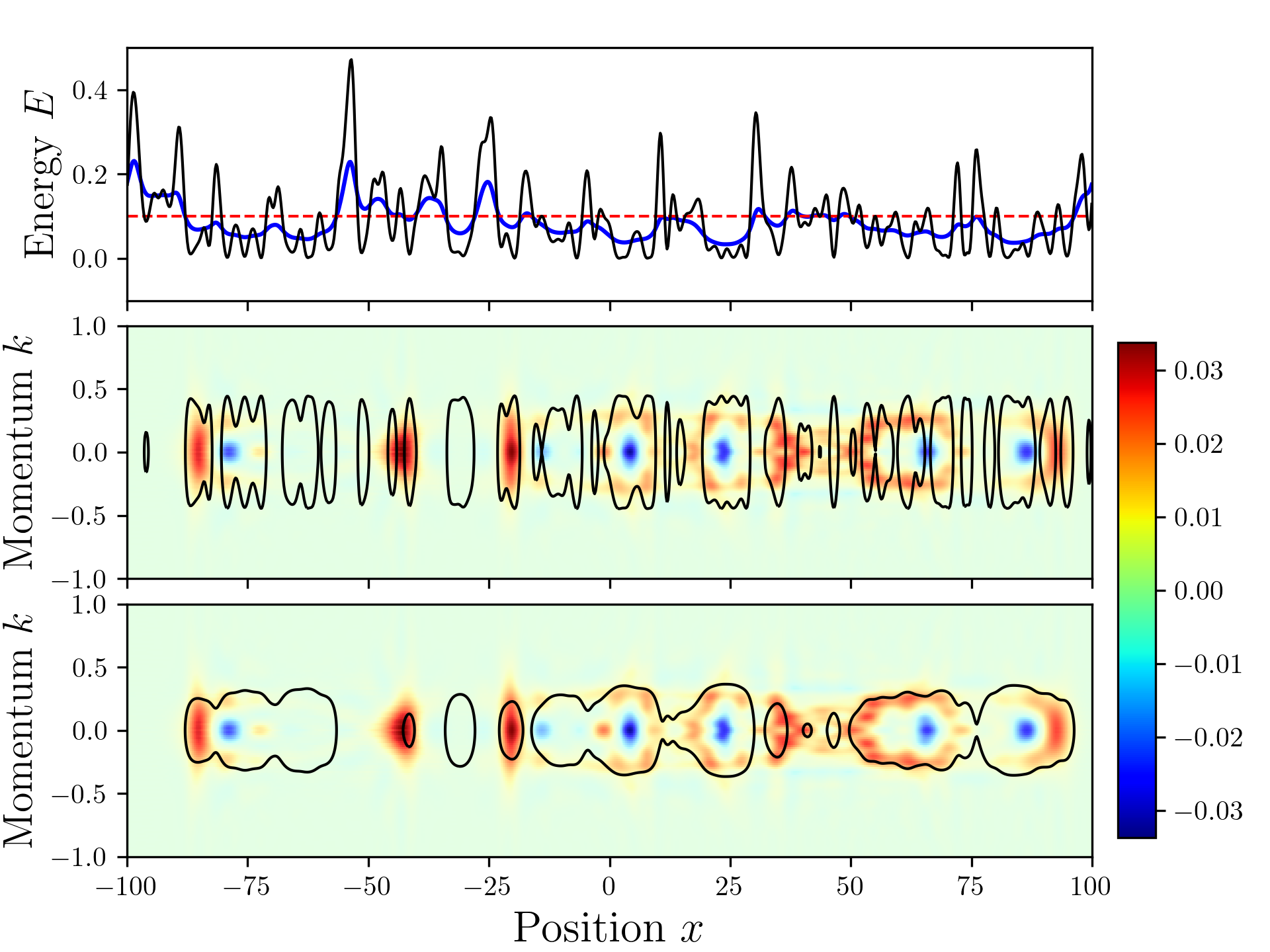}
\caption{\textbf{Deep quantum regime}. Top frame: disordered potential $V$ with $\eta=0.1$ (black line), effective potential $V_u$ (blue line) and energy level at $E=0.1$ (dashed red line). Middle frame: color plot of $F_E(x,k)$ defined in Eq.~\eqref{eq:FE} computed over all eigenfunctions in a range $\Delta E = 0.2 E$ around the target energy $E=0.1$, superimposed with the level set $H(x,k)=E$ represented by a continuous black line. Bottom frame: Similar representation as in the middle frame, but with the level set $H_1(x,k)=E$.}
\label{fig:eta01E01}
\end{figure}

For each figure (\ref{fig:eta2} to \ref{fig:eta01E01}), the top frame displays the original potential~$V$ and the effective potential $V_u = 1/u$. For the middle frame, we average the Wigner functions of the $N_{[E-\Delta E,E+\Delta E]}$ eigenstates lying in the $[E-\Delta E,E+\Delta E]$ energy interval
\begin{equation}\label{eq:FE}
F_E(x,k) = \frac{1}{N_{[E-\Delta E,E+\Delta E]}} \sum_{i} W_i(x,k) \,,
\end{equation}
and superimpose the color representation of the contours of $F_E(x,k)$ with the level set of $H(x,k)=E$. The bottom frame displays the contours of $F_E(x,k)$, but this time superimposed with the level set $H_1(x,k)=E$.

Figure~\ref{fig:eta2} shows the results for a disorder strength~$\eta=2.0$ around a target energy~$E=0.5$. In the top frame, the energy level (dashed red line) crosses the graph of the effective potential (blue line) in its lower part, meaning that the quantum states are well localized within the wells predicted by the effective potential. In the middle frame, one sees that the level set of the original Hamiltonian $H(x,k)$ at energy~$E$ (black line) presents a large number of closed curves, many of which that do not correspond to any localized eigenfunction around~$E$ (materialized by $F_E(x,k)$). On the contrary, in the bottom frame, the location of the averaged Wigner function $F_E(x,k)$ in phase space appears to be  much more accurately determined by the contours of $H_1(x, k)$. Empty contours correspond to states having an energy lower than $E-\Delta E$, not included in the average~\eqref{eq:FE}.

We now move to a more quantum regime ($\eta=0.5$) and explore an energy range around the average potential value ($E=0.5$), in order to observe excited states (see Fig.~\ref{fig:eta05}). Here again, the level set of $H_1(x,k)=E$ follows the contours of the distribution $F_E(x,k)$ much better than the one of the original potential. 

In the same regime ($\eta=0.5$), Fig.~\ref{fig:eta05E2} displays states at a higher energy ($E=2$) compared to the potential strength. In this case, the Wigner functions are weakly affected by the disorder and lie around the two horizontal lines of equation $k^2/2=E$, which in this case amounts to $k = \pm 2$. The level sets of both $H_1(x,k)$ and $H(x,k)$ delineate the support of the distribution $F_E(x,k)$, although $H(x,k)$ exhibits larger short range fluctuations which are not present in the Wigner functions.

Finally, we explore a deep quantum regime ($\eta=0.1$) in Figures~\ref{fig:eta01lowest} and \ref{fig:eta01E01}. At the bottom of the spectrum (Fig.~\ref{fig:eta01lowest}), it is very clear now that the structure of $F_E(x,k)$ is much more accurately predicted by the level sets of $H_1$. In particular, in the middle frame, one sees that the level set $H(x,k)=E$ encloses a large number of elongated regions where the Wigner function remains very small. Such regions do not appear anymore inside the level set $H_1(x,k)=E$. Note that one observes spurious oscillations in the level curves of the original Hamiltonian which do not reflect the actual behavior of the Wigner functions. These oscillations are not found in the effective Hamiltonian. This observation still holds at higher energy ($E=0.1$, see Fig.~\ref{fig:eta01E01}).

The conclusion of this section is that, across all energy ranges and across different regimes (corresponding to different values of the parameter $\eta$), the level sets of $H_1$ reproduce the main features of the Wigner functions. This agreement supports and explains the accuracy of the approximation~\eqref{eq:N_Vu}. Moreover, it provides a more accurate way to assess how the energy is actually distributed in phase space than the original Hamiltonian, i.e., $H(x,k)$. We will now see how $H_1(x,k)$ can help us compute spectral functions in laser speckle potentials, both in classical and quantum regimes, without any adjustable parameters.

\section{The spectral function estimates}

The spectral function $A_{\vb{k}}(E)$ can be understood as the ensemble average of the energy distribution of a plane wave of wave vector $\vb{k}$. It is mathematically defined as
\begin{align}\label{eq:spectralfunction}
    A_{\vb{k}}(E) &\equiv \overline{\sum_\alpha \abs{\braket{\vb{k}}{\varphi_\alpha}}^2 \, \delta(E- E_\alpha)} \\
    &= \overline{\bra{\vb{k}}\delta(E-\hat{H})\ket{\vb{k}}} \,,\nonumber
\end{align}
where the sum is taken over all eigenfunctions $\varphi_\alpha$ of the Hamiltonian having an energy $E_\alpha = E$. It can also be physically interpreted as the ensemble average momentum distribution of states of energy $E$. More generally, one can extend this definition to the ensemble average of the energy distribution of any quantum state~$\ket{\psi}$:
\begin{equation}\label{eq:genspectralfunction}
    \mathcal{A}_\psi(E) = \overline{\sum_\alpha \abs{ \braket{\psi}{\varphi_\alpha} }^2 \delta(E- E_\alpha)} \,.
\end{equation}

The spectral function is directly related to the average retarded Green function of the system. Indeed, using the identity
\begin{equation}
	 \delta(E-\hat{H}) = \lim_{\eta\to 0^+} - \frac{1}{\pi} \Im\left( \frac{1}{E-\hat{H}+i\eta} \right)\,,
\end{equation}	 
a time-energy Fourier transform makes it possible to express the spectral function as
\begin{equation}
A_{\vb{k}}(E) =  \frac{1}{\pi} \Re \int_0^{\infty}{\overline{\bra{\vb{k}}\mathrm{e}^{-i\hat{H}t}\ket{\vb{k}}}\ \mathrm{e}^{iEt}\ dt} \,.
\end{equation}
The spectral function can thus be numerically computed by following the time evolution of an initial plane wave~$\ket{\vb{k}}$ governed by the Hamiltonian $\hat{H}$. This computation is performed using an expansion of the evolution operator in polynomials of the Hamiltonian, see~\cite{Trappe2015} for details. One should note that, in the semiclassical limit, the spectral function at $\vb{k}=\vb{0}$ tends to the distribution of the values of the original potential~$V$. We will see in the following that the LL-based approach provides a way to generalize this statement even in the deep quantum regime.

\subsection{The LL-based Wigner-Weyl approach}

The Wigner-Weyl theory, and specifically Eq.~\eqref{eq:expval}, tell us that the average energy when the system is in the state $\ket{\psi}$ is
\begin{equation}\label{eq:exactWeyl}
    \bra{\psi}\hat{H}\ket{\psi} = \iint W_\psi(\vb{x}, \vb{k}) \, H(\vb{x}, \vb{k}) \, \mathrm{d}\vb{x} \, \mathrm{d}\vb{k} \,.
\end{equation}
One can formally rewrite this identity by grouping together all points $(\vb{x}, \vb{k})$ in phase space that correspond to the same value of the classical energy
\begin{align}\label{eq:gpsi}
\bra{\psi}\hat{H}\ket{\psi} &= \int E \left( \iint_{E \leq H(\vb{x}, \vb{k}) \leq E+dE} W_\psi(\vb{x}, \vb{k}) \, \mathrm{d}\vb{x} \, \mathrm{d}\vb{k} \right) \nonumber\\
 &= \int E  ~g_\psi(E)\, dE \,.
\end{align}
The function $g_\psi(E)$ could be interpreted as the energy distribution function of $\psi$, i.e., the function~$\mathcal{A}_\psi(E)$ defined in~\eqref{eq:genspectralfunction}. In other words, $g_\psi(E) \, dE$ could be understood as the norm of the projection of~$\psi$ on the eigenstates of energy comprised in $[E,E+dE]$. However, we have seen in the previous section that the supports of the eigenstates of the Hamiltonian are more precisely enclosed by the level sets of $H_1(\vb{x},\vb{k})$. It is natural therefore to replace $H(\vb{x}, \vb{k})$ by $H_1(\vb{x}, \vb{k})$ for approximating the energy distribution, yielding
\begin{equation}\label{eq:conjecture-Apsi}
\mathcal{A}_\psi(E) \, dE \approx \iint_{\scriptsize \begin{array}{c} E \leq H_1(\vb{x}, \vb{k}) \\ \leq E + dE\end{array}} W_\psi(\vb{x}, \vb{k}) \, \mathrm{d}\vb{x} \, \mathrm{d}\vb{k} \,.
\end{equation}
The specific case of plane waves ($\ket{\psi} = \ket{\vb{k}}$) would lead to an estimate on the spectral function. It requires the Wigner function of a plane wave. For normalization reasons, one has to restrain the plane wave to a bounded spatial domain $\Omega$ of volume~$\abs{\Omega}$. A straightforward computation 
leads to
\begin{align}
    W_{\vb{k_0}}(\vb{x}, \vb{k}) &= \frac{1}{(2\pi)^d \abs{\Omega}} \int e^{-i \vb{k_0} \cdot 
(\vb{x} - \frac{\vb{y}}{2})} e^{i \vb{k_0} \cdot (\vb{x} + \frac{\vb{y}}{2})} e^{-i \vb{k} \cdot \vb{y}} \mathrm{d} \vb{y} \nonumber\\
    &= \frac{1}{\abs{\Omega}} ~\delta(\vb{k}-\vb{k_0}) \,.
\end{align}
We see that, not surprisingly, the Wigner function of a plane wave of wave vector $\vb{k_0}$ is simply a normalized delta-distribution along $k$. Inserting this into Eq.~\eqref{eq:conjecture-Apsi} gives immediately a conjectural estimate for the spectral function:
\begin{align}\label{eq:conjecture-spectral}
    A_{\vb{k_0}}(E) \,dE &\approx \frac{1}{\abs{\Omega}} \overline{\left (\iint_{\scriptsize \begin{array}{c} E \leq H_1(\vb{x}, \vb{k}) \\ \leq E + dE\end{array}} \delta(\vb{k}-\vb{k_0})\, \mathrm{d}\vb{k} \, \mathrm{d}\vb{x} \right)} \nonumber \\
    &= \frac{1}{\abs{\Omega}} \overline{\left ( \int_{E-\frac{\vb{k_0}^2}{2} \leq V_u(\vb{x}) \leq E - \frac{\vb{k_0}^2}{2} + dE} \mathrm{d} \vb{x}  \right)} \,.
\end{align}
This last quantity is the normalized distribution of the values taken by $V_u$ (in other words, the  histogram of the values of $V_u$) shifted by a $\vb{k_0}^2/2$:
\begin{align}
A_{\vb{k_0}}(E) = \mathcal{P}\left(E - \frac{\vb{k}_0^2}{2}\right) \,,
\end{align}
where $\mathcal{P}$ is the probability density of $V_u$. In particular, in the case $\vb{k_0} = \vb{0}$, the spectral function takes a very simple form. It is nothing but the normalized distribution of the values taken by the effective potential~$V_u$:
\begin{align}\label{eq:conjecture-AkE}
A_{\vb{0}}(E) = \mathcal{P}\left(E \right) \,.
\end{align}
This result shows us how the LL-based approach generalizes the understanding of the semiclassical limit of the spectral function for $\vb{k}=0$, even in the quantum regime. The effective potential~$V_u$ automatically incorporates the quantum effects such as confinement or tunneling, and hence performs an automatic renormalization of the disorder. The distribution of values of $V_u$ replaces here the distribution of values of~$V$, thus providing a very good approximation of the spectral function for any value of~$\eta$. In the semiclassical limit, $V_u$ is very similar to $V$ and we recover the known expression of the spectral function. We will now test this conjectural expression for several types of disorder and disorder strengths.

\subsection{Numerical computations}

\begin{figure}[ht!]
\centering
\includegraphics[width=.49\columnwidth]{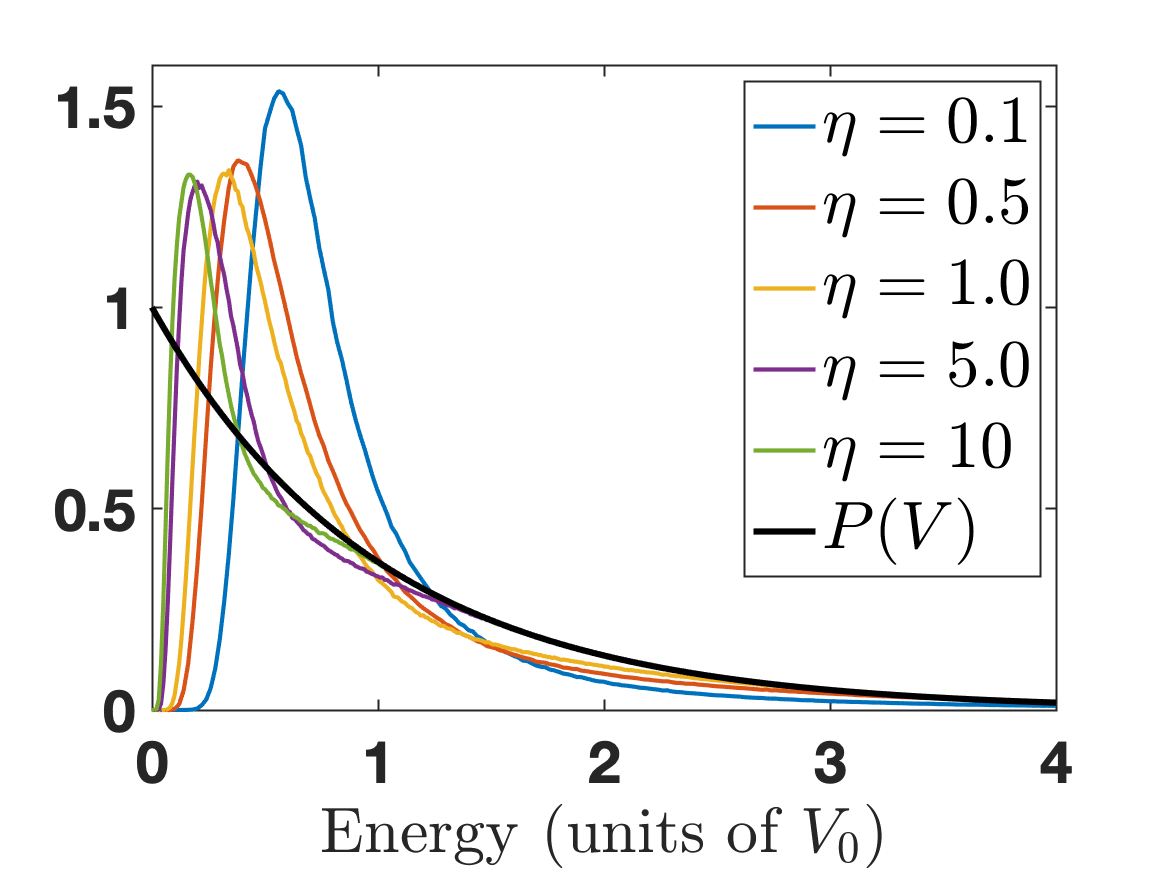}
\includegraphics[width=.49\columnwidth]{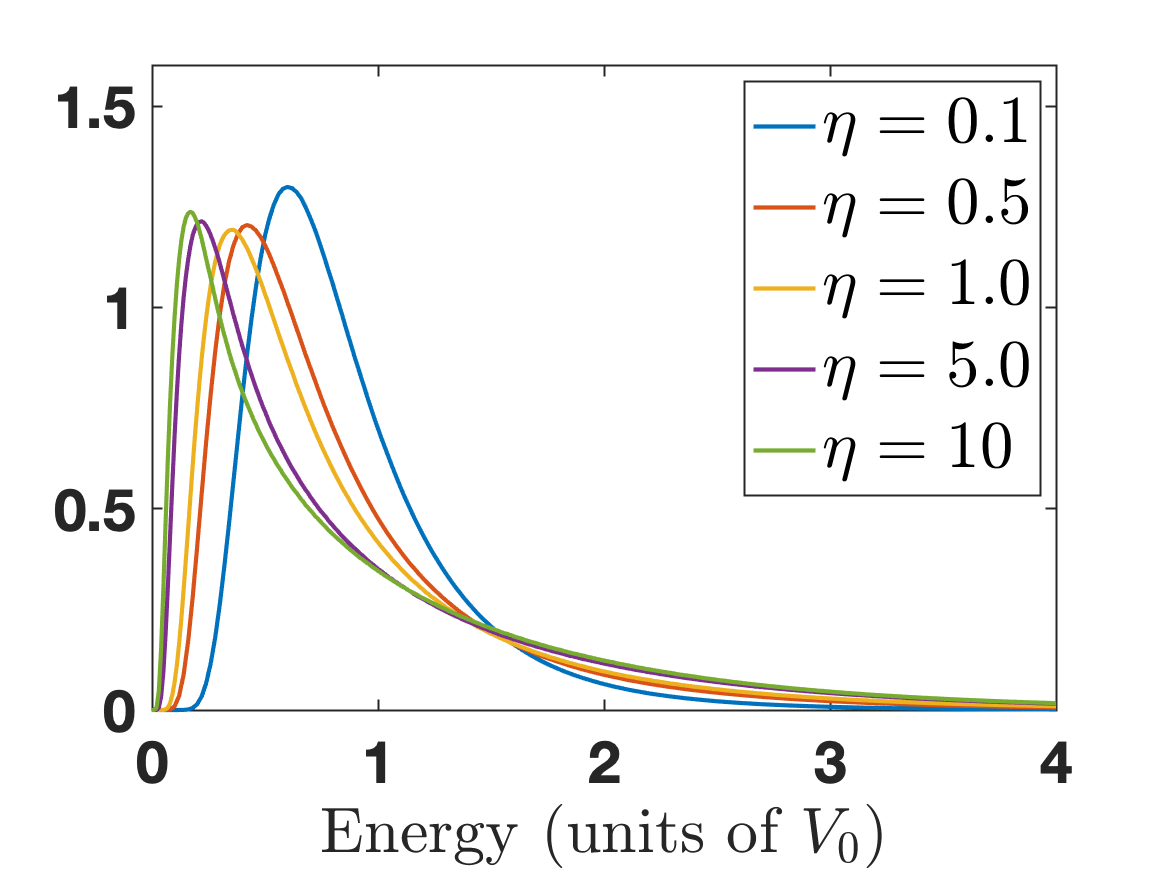}\\
\includegraphics[width=.49\columnwidth]{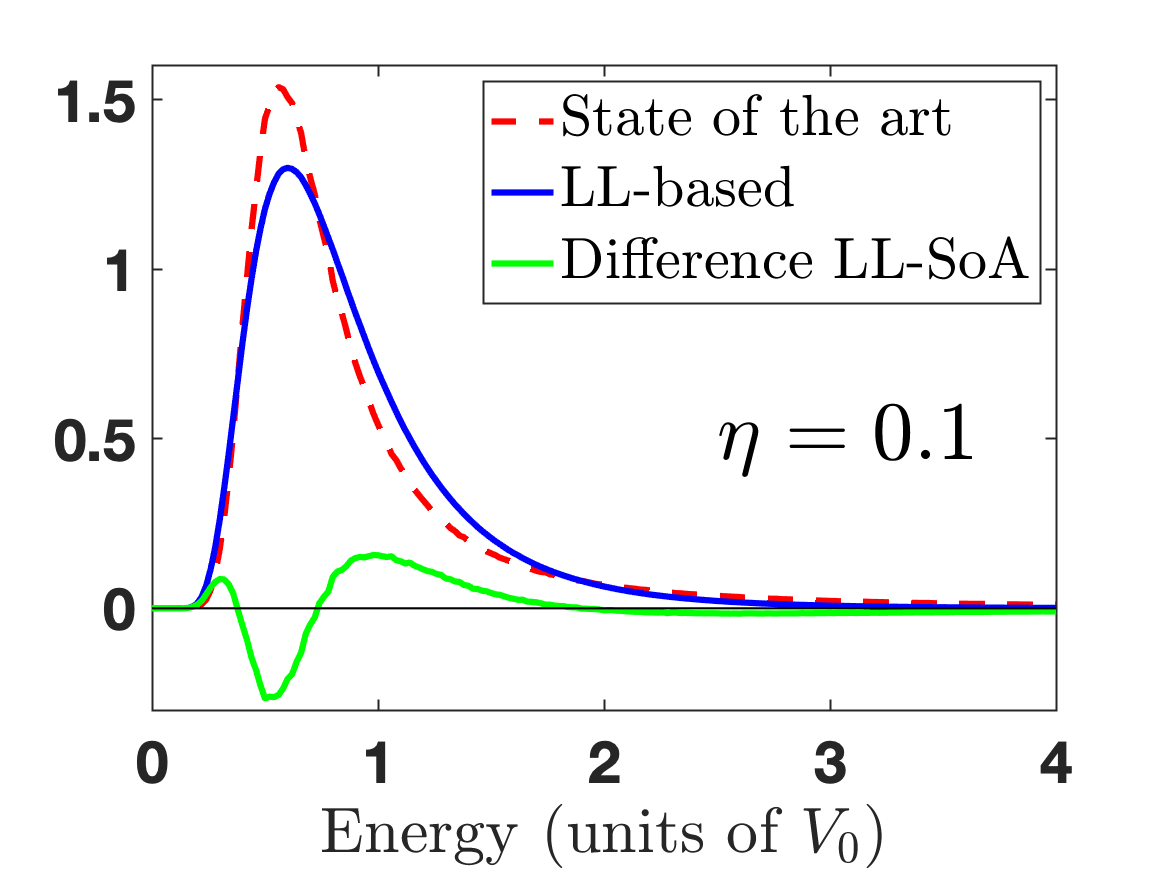}
\includegraphics[width=.49\columnwidth]{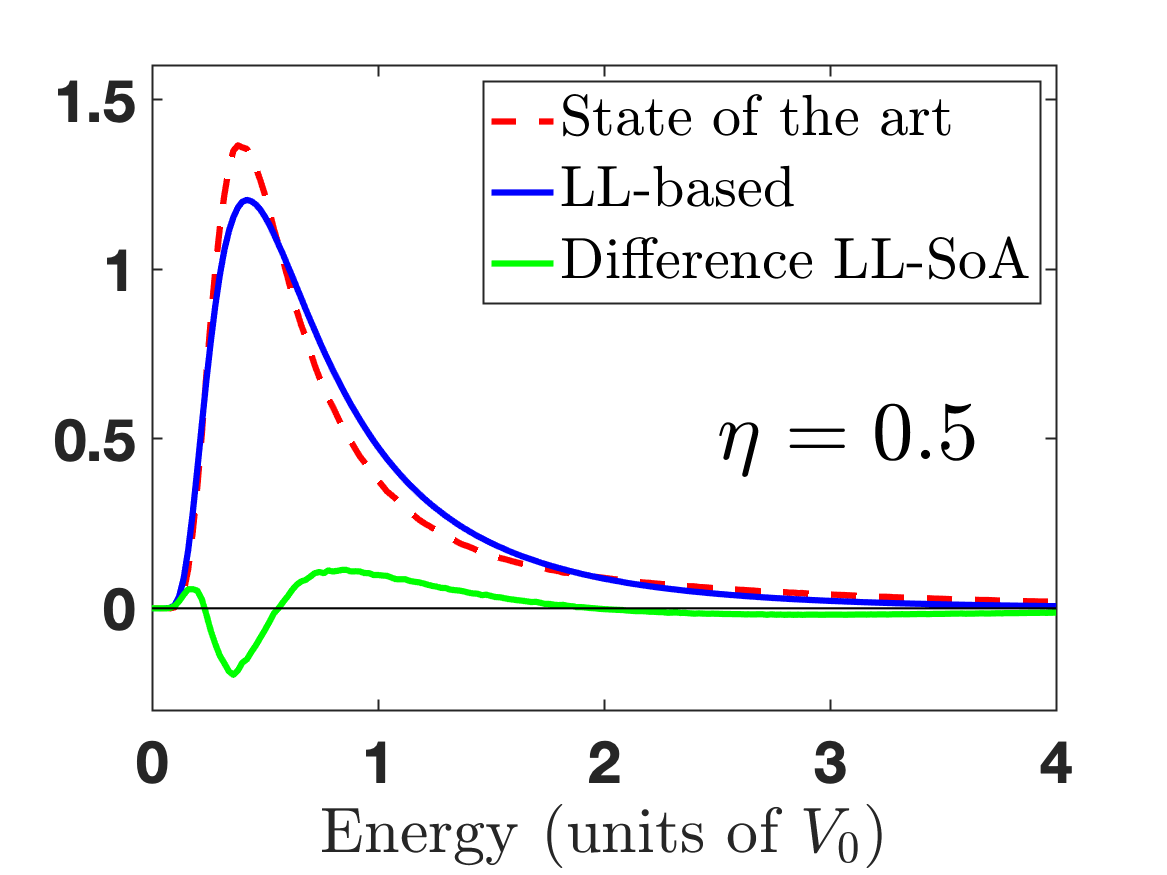}\\
\includegraphics[width=.49\columnwidth]{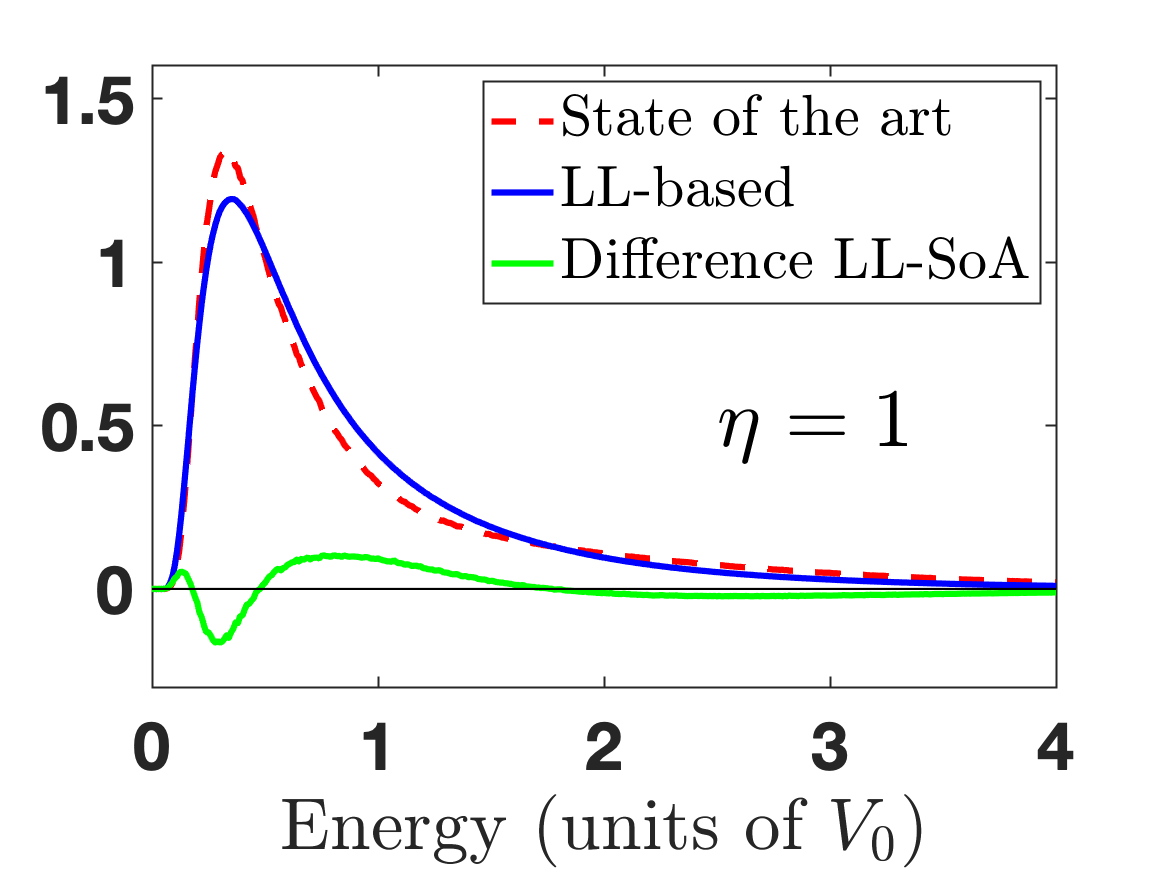}
\includegraphics[width=.49\columnwidth]{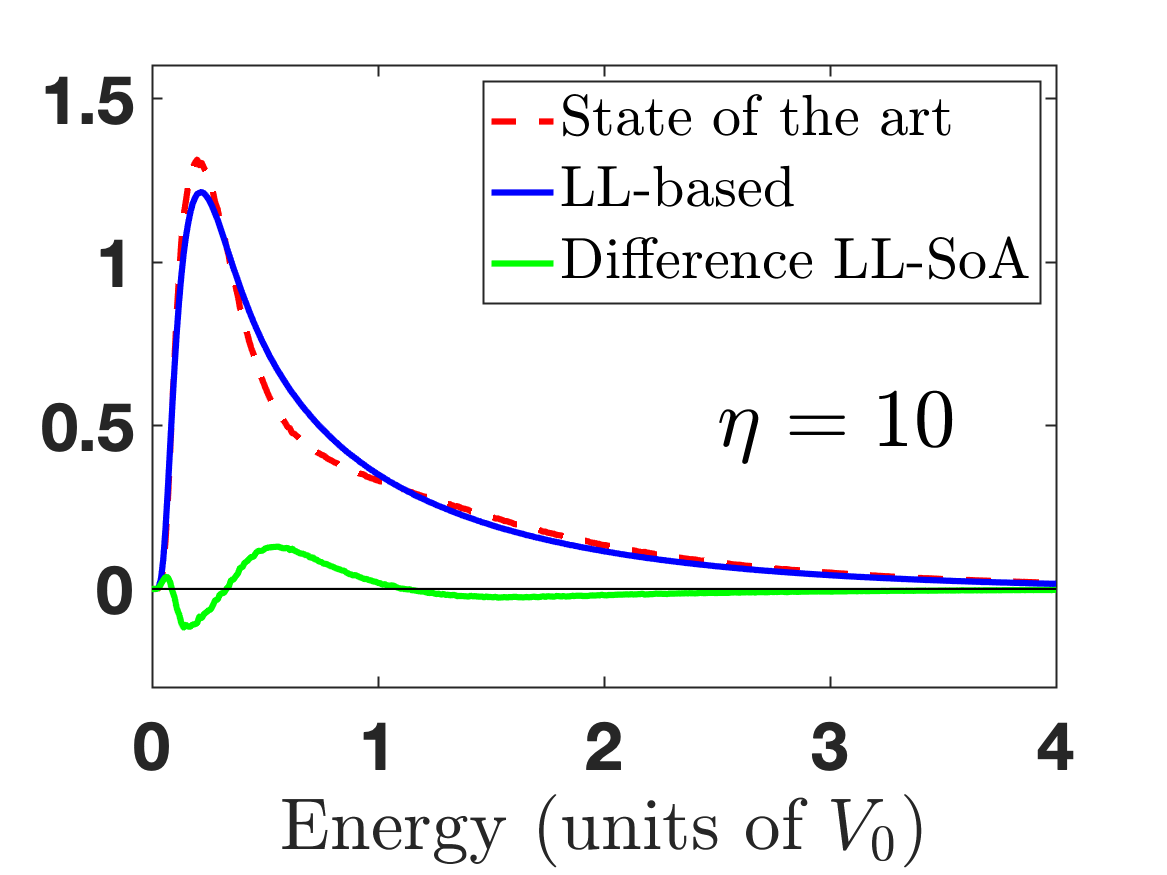}
\caption{Top left: Spectral functions (in units of $1/V_0$) in the case of a Gaussian-correlated speckle potential, computed from the solutions to the Schrödinger equation for 5~different values of the parameter $\eta$ ranging from 0.1  to 10 (0.1, 0.5, 1, 5, and 10). The distribution of the values of the original potential~$V$ is also plotted in black. Top right: Spectral functions computed from the LL theory for the same values of~$\eta$. Middle and bottom panels: Direct comparisons between the spectral functions computed using the Schrödinger equation (dashed red) and the LL theory (blue), for $\eta=0.1$, 0.5, 1, and 10, respectively. The difference between both curves (the error) is plotted in green.}
\label{fig:1DspeckleGaussian}
\end{figure}

We first test our conjectural expression~\eqref{eq:conjecture-AkE} on blue-detuned speckle potentials with Gaussian correlation function, whose characteristics are given in \eqref{eq:speckleGaussian}. The system size is $L=300$ and the discretization step is $\delta = 0.05$. For each value of the parameter $\eta$, the ensemble average of Eq.~\eqref{eq:conjecture-spectral} is achieved over 50,000 realizations of the potential. This LL-based estimate is compared to state-of-the-art exact numerical computations of the spectral functions~\cite{Trappe2015}. The two top frames of Fig.~\ref{fig:1DspeckleGaussian} display state-of-the-art computations (left frame) and LL-based formulas (right frame) for 5 differents values of $\eta$. Each of the following four frames of Fig.~\ref{fig:1DspeckleGaussian} corresponds to a different value of $\eta$ (hence of $V_0$, see Eq.~\eqref{eq:eta}), from $\eta=0.1$ (top left, deep quantum regime) to $\eta=10$ (bottom right, semiclassical regime). In each frame, three curves are plotted: the exact calculation of the ensemble-averaged spectral function for $k_0=0$ (red dashed line), the LL-based estimate using Eq.~\eqref{eq:conjecture-spectral} (blue line), and the difference between the latter and the former quantities (green line).

These results lead to several observations. First, the spectral functions at various values of $\eta$ are quite different, evolving from a close-to-symmetric shape at small~$\eta$ to a very asymmetric function at large $\eta$, where it is close to the disorder distribution given in Eq.~\eqref{eq:speckleGaussian}. Secondly, the LL-based estimates reproduce the behavior of the spectral function, both in the semiclassical and in the quantum regime, without any adjustable parameter. In particular, the positions of the peak and the widths of the distribution are almost identical. Thirdly, the difference between the estimate and the spectral function (i.e., the error plotted in green) displays a similar pattern in all regimes, suggesting that the LL-based estimate could be the first term of an expansion.

\begin{figure}[ht!]
\centering
\includegraphics[width=.49\columnwidth]{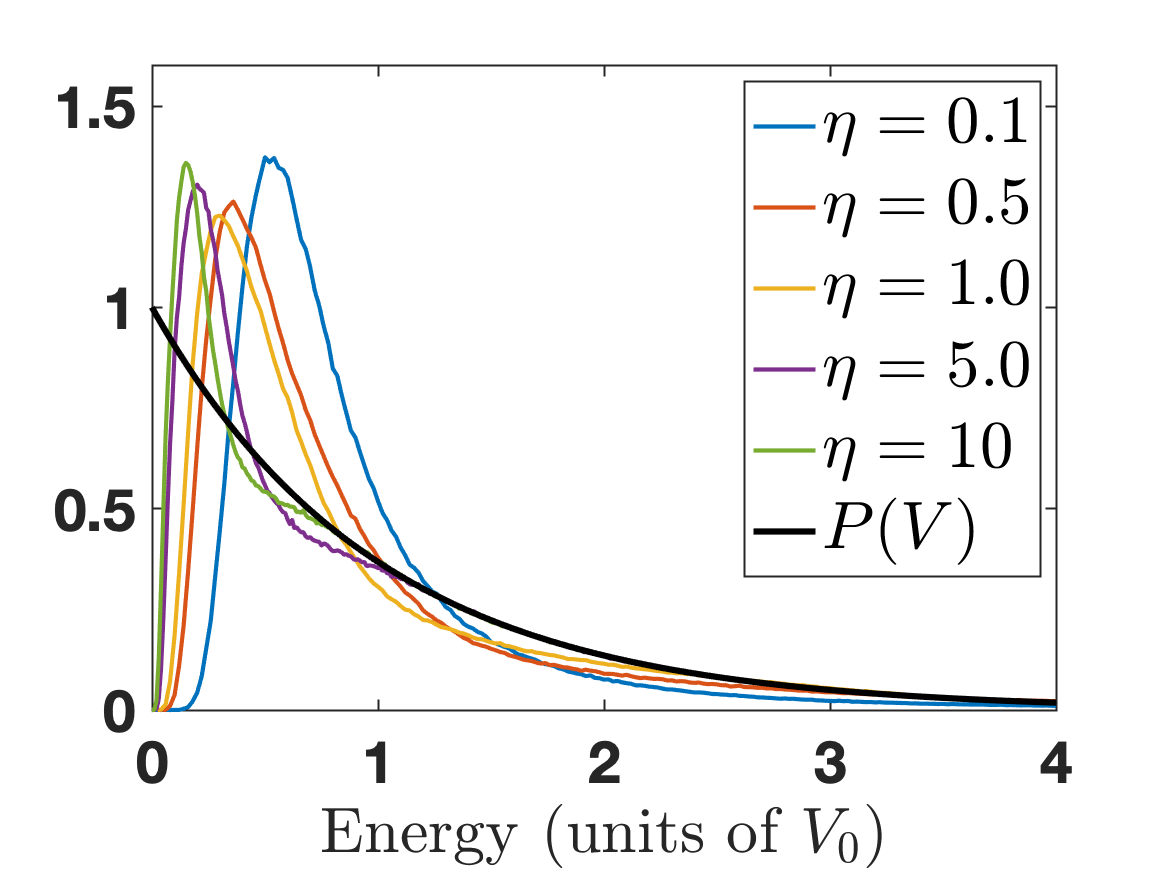}
\includegraphics[width=.49\columnwidth]{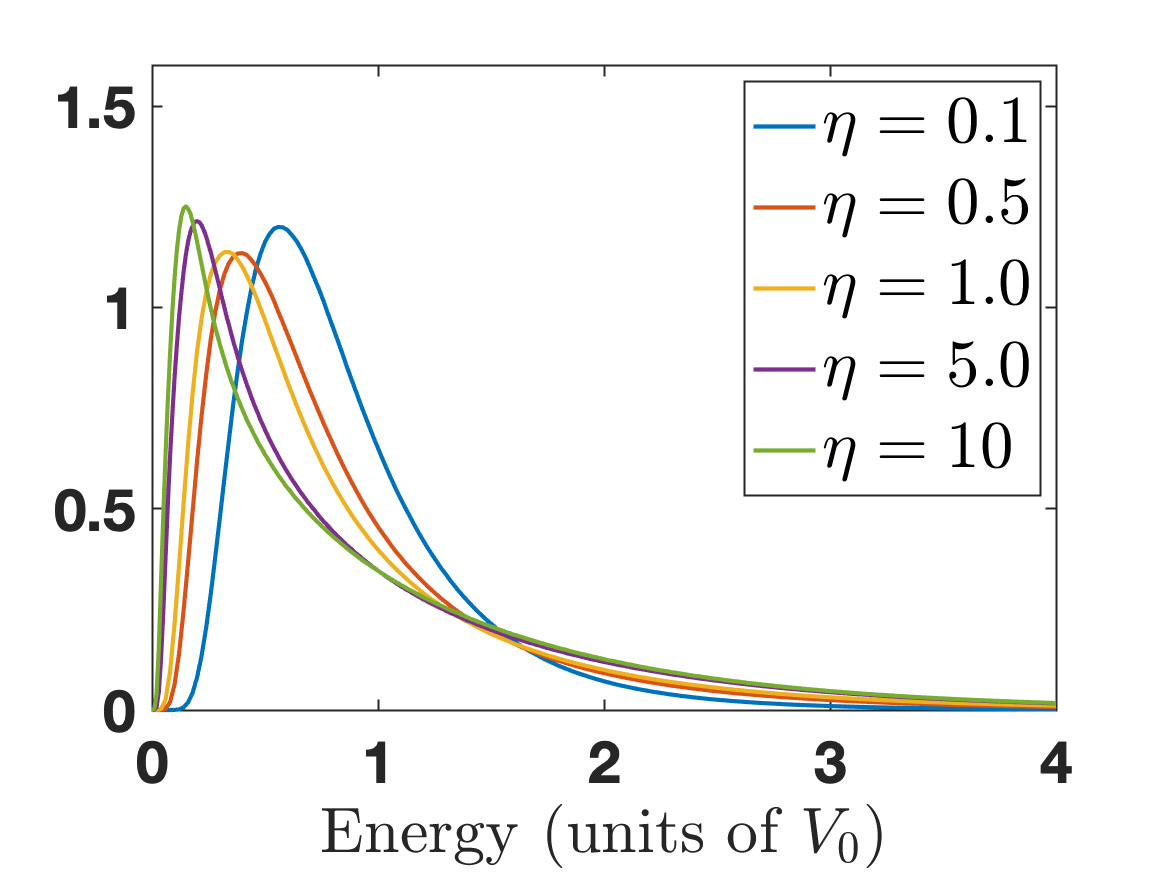}\\
\includegraphics[width=.49\columnwidth]{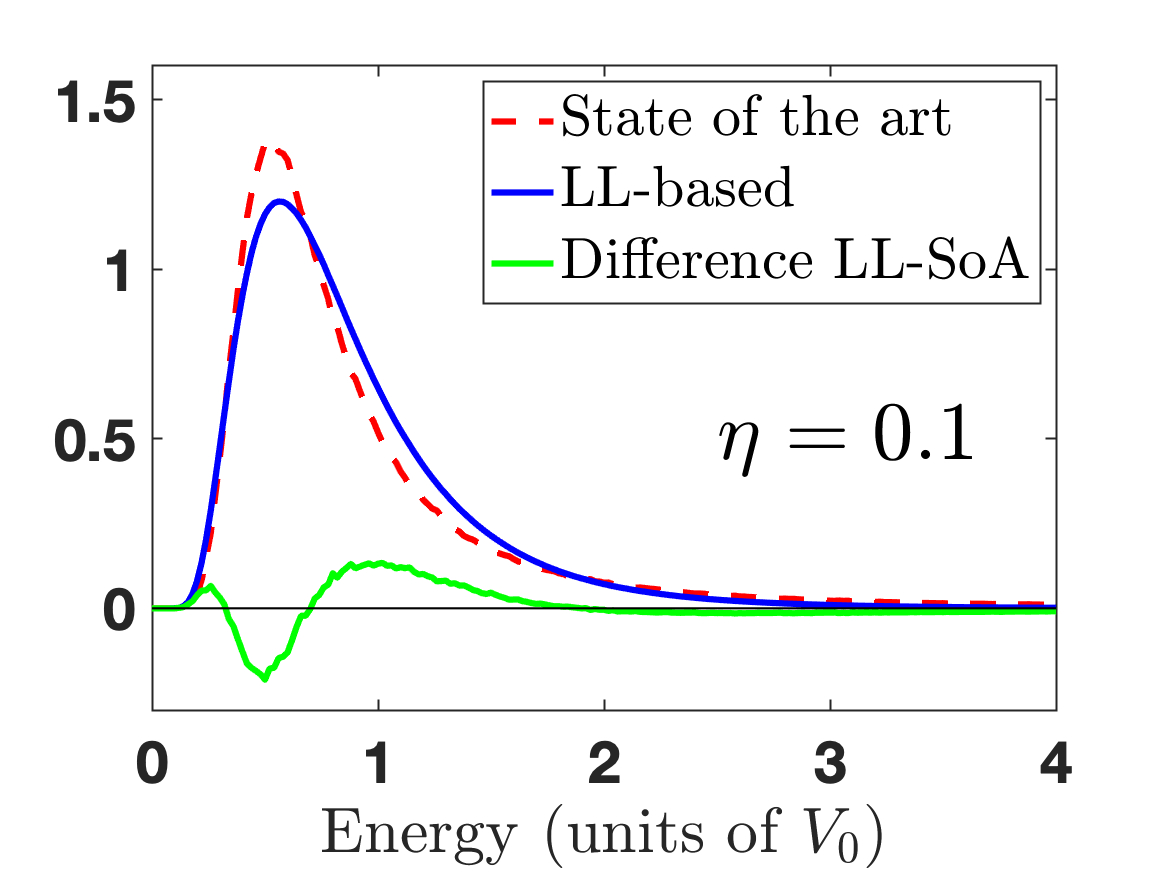}
\includegraphics[width=.49\columnwidth]{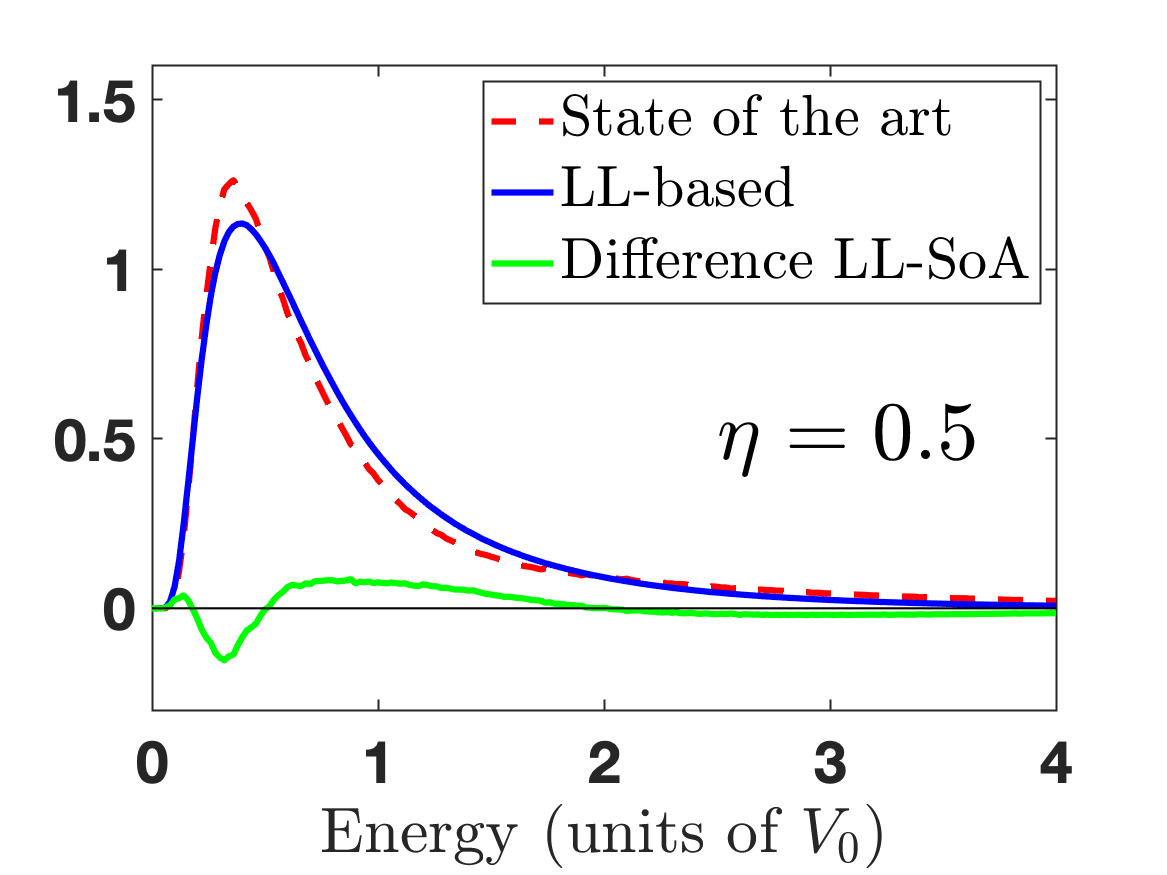}\\
\includegraphics[width=.49\columnwidth]{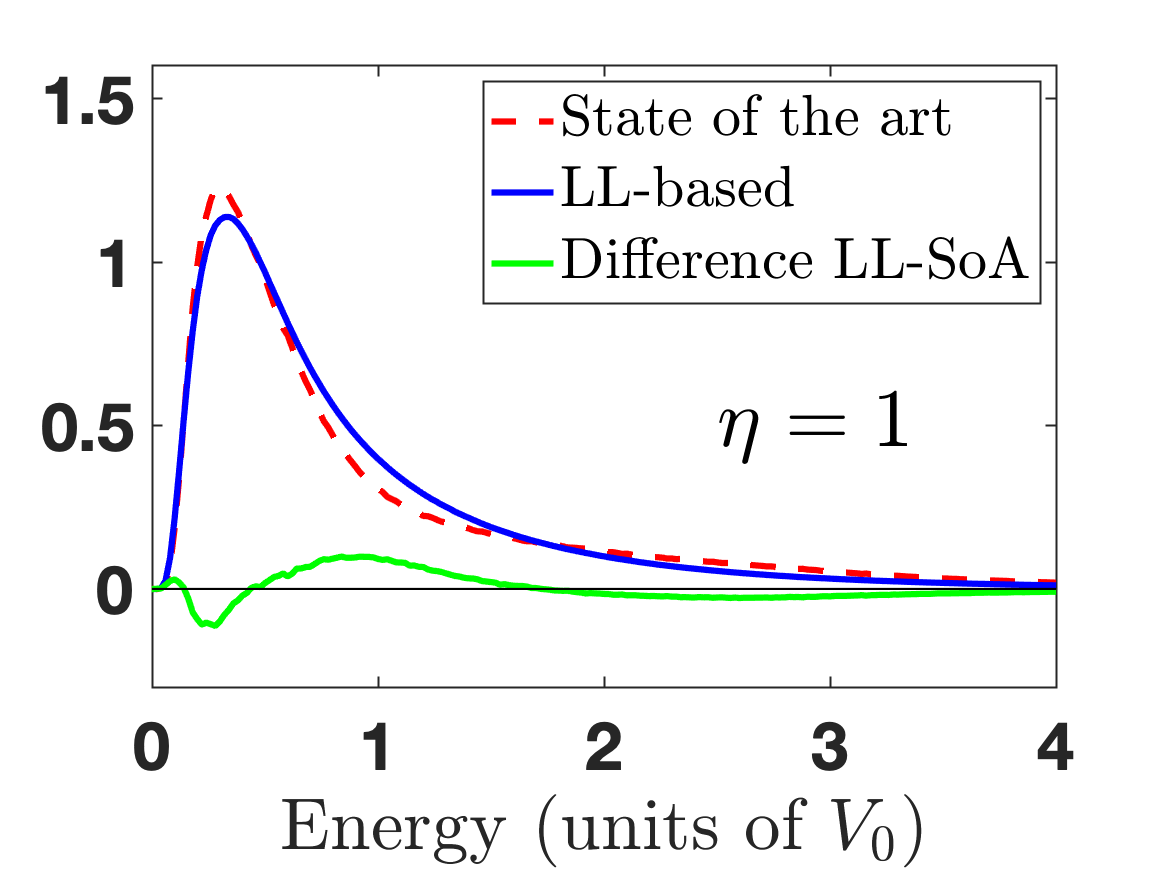}
\includegraphics[width=.49\columnwidth]{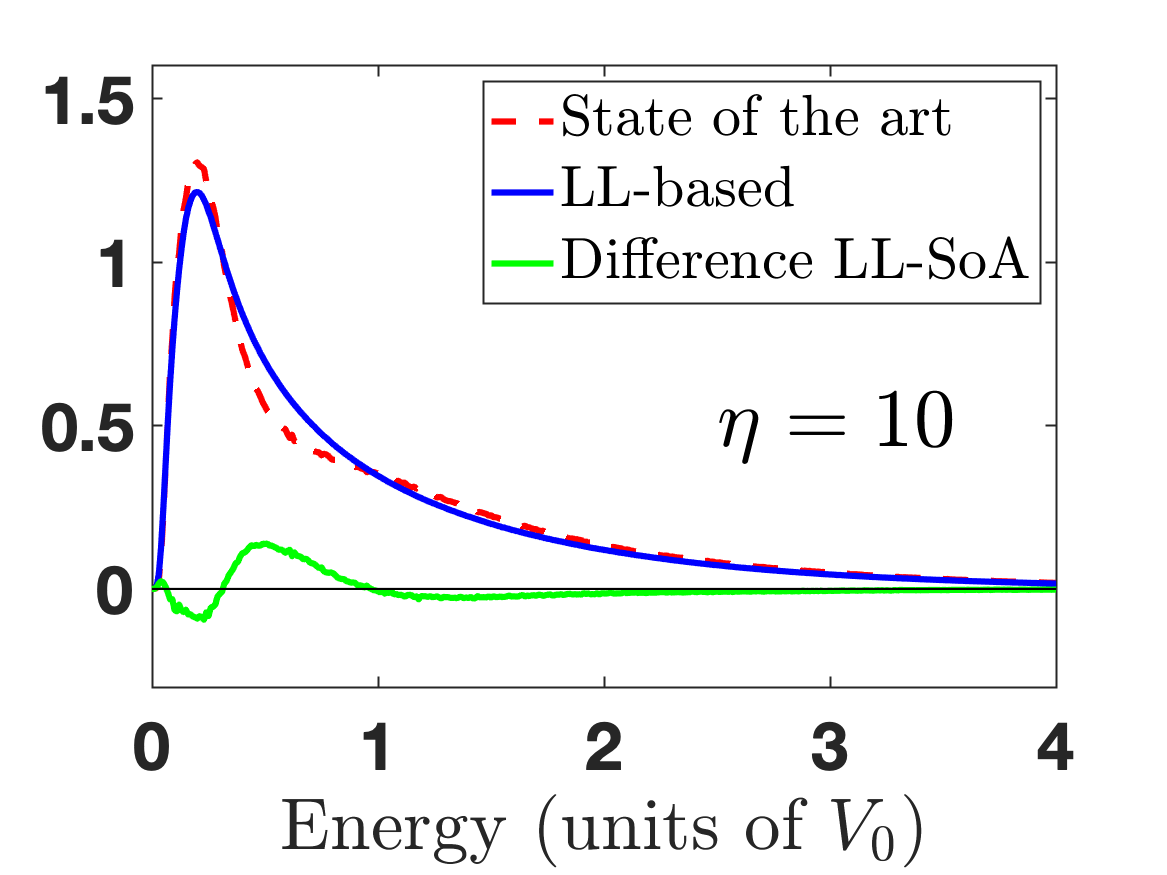}
\caption{Top left: Spectral functions (in units of $1/V_0$) in the case of a sinc-correlated speckle potential, computed from the solutions to the Schrödinger equation for 5~different values of the parameter $\eta$ ranging from 0.1  to 10 (0.1, 0.5, 1, 5, and 10). The distribution of the values of the original potential~$V$ is also plotted in black. Top right: Spectral functions computed from the LL theory for the same values of $\eta$. Middle and bottom panels: Direct comparisons between the spectral functions computed using the Schrödinger equation (dashed red) and the LL theory (blue), for $\eta=0.1$, 0.5, 1, and 10, respectively. The difference between both curves (the error) is plotted in green.}
\label{fig:1Dspecklesinc}
\end{figure}

We then tested our approach for a different potential, characterized by the same exponential distribution of values, but with a ``sinc-type'' correlation function, corresponding to the specke created by a slit~\cite{Billy2008}:
\begin{equation}\label{eq:sinc-corr}
    g(x) = V_0^2 \, \textrm{sinc}^2 \left(\frac{x}{\sigma} \right) \,.
\end{equation}
The results, shown in Fig.~\ref{fig:1Dspecklesinc}, are averaged over 10,000~realizations. The plots are remarkably similar to those in Fig.~\ref{fig:1DspeckleGaussian} as the spectral functions depend very weakly on the shape of the correlation function.

\begin{figure}[ht!]
\centering
\includegraphics[width=.49\columnwidth]{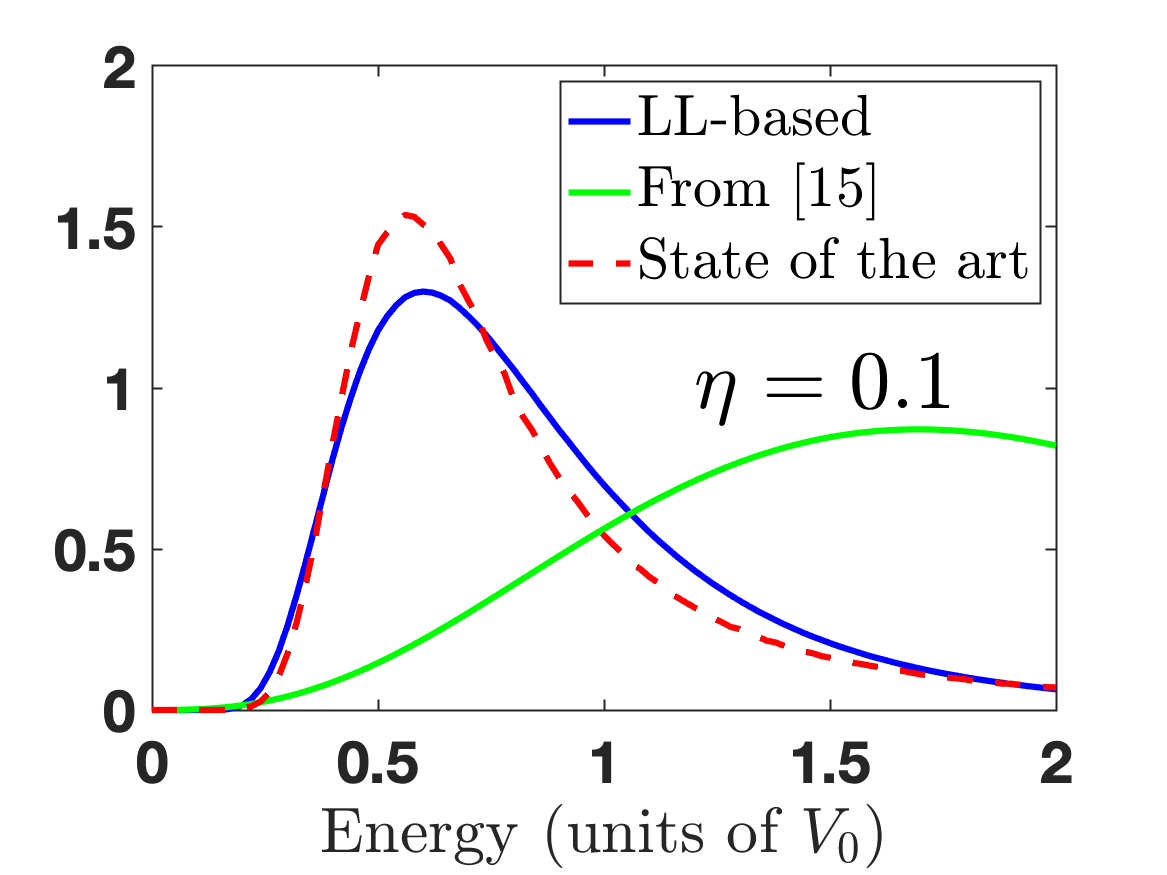}
\includegraphics[width=.49\columnwidth]{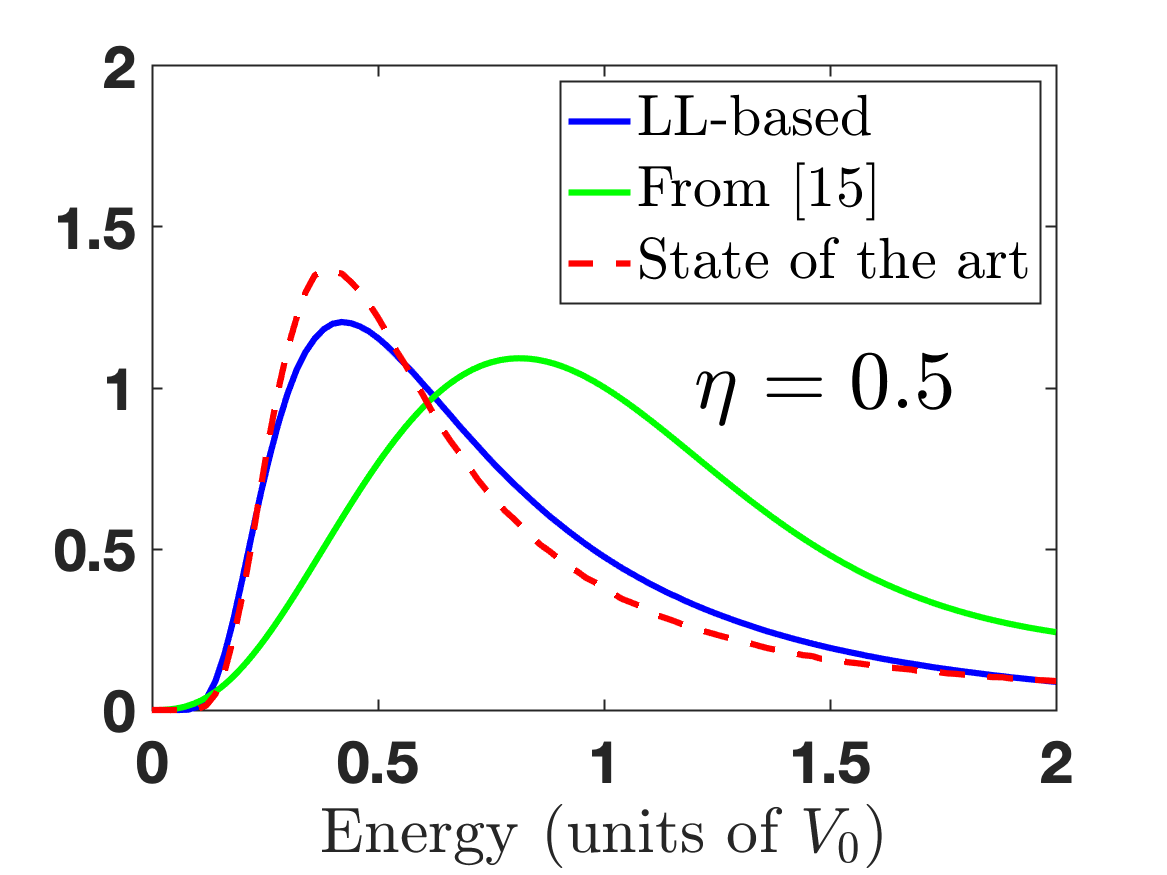}\\
\includegraphics[width=.49\columnwidth]{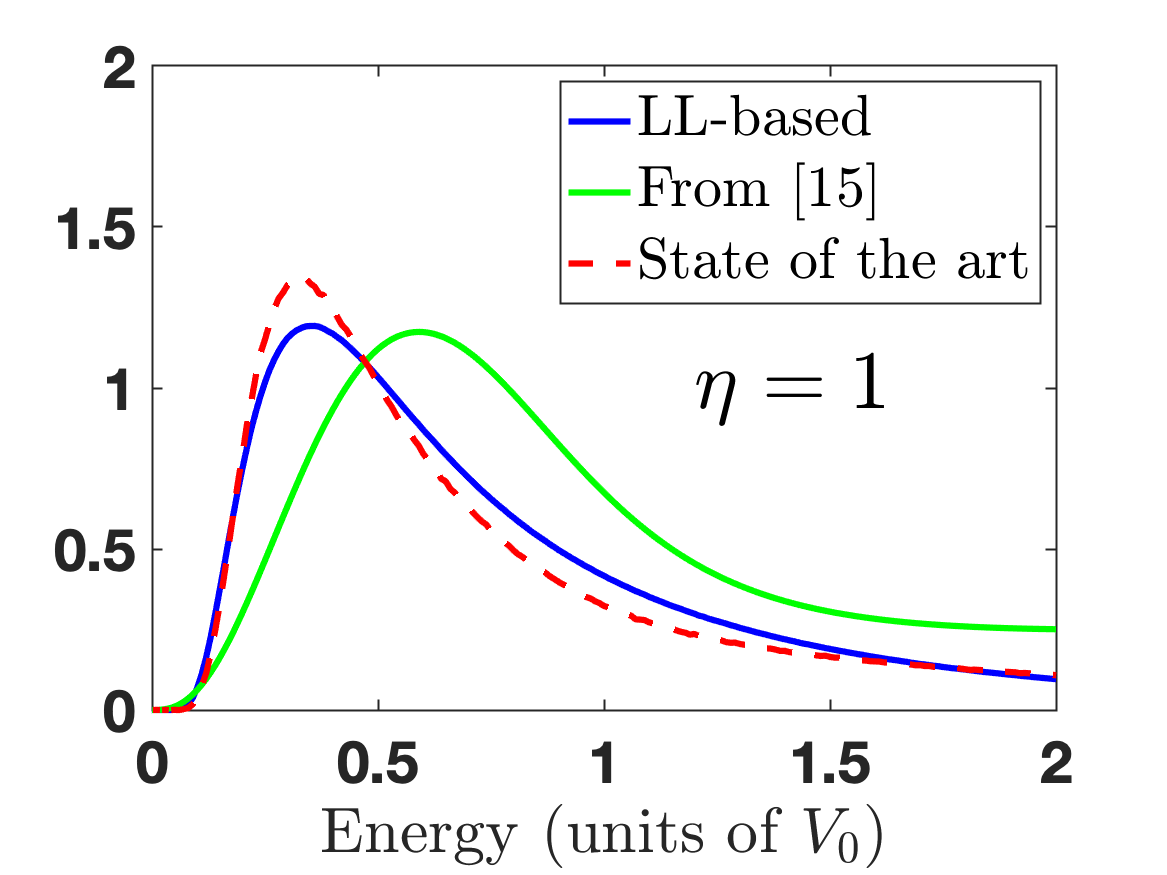}
\includegraphics[width=.49\columnwidth]{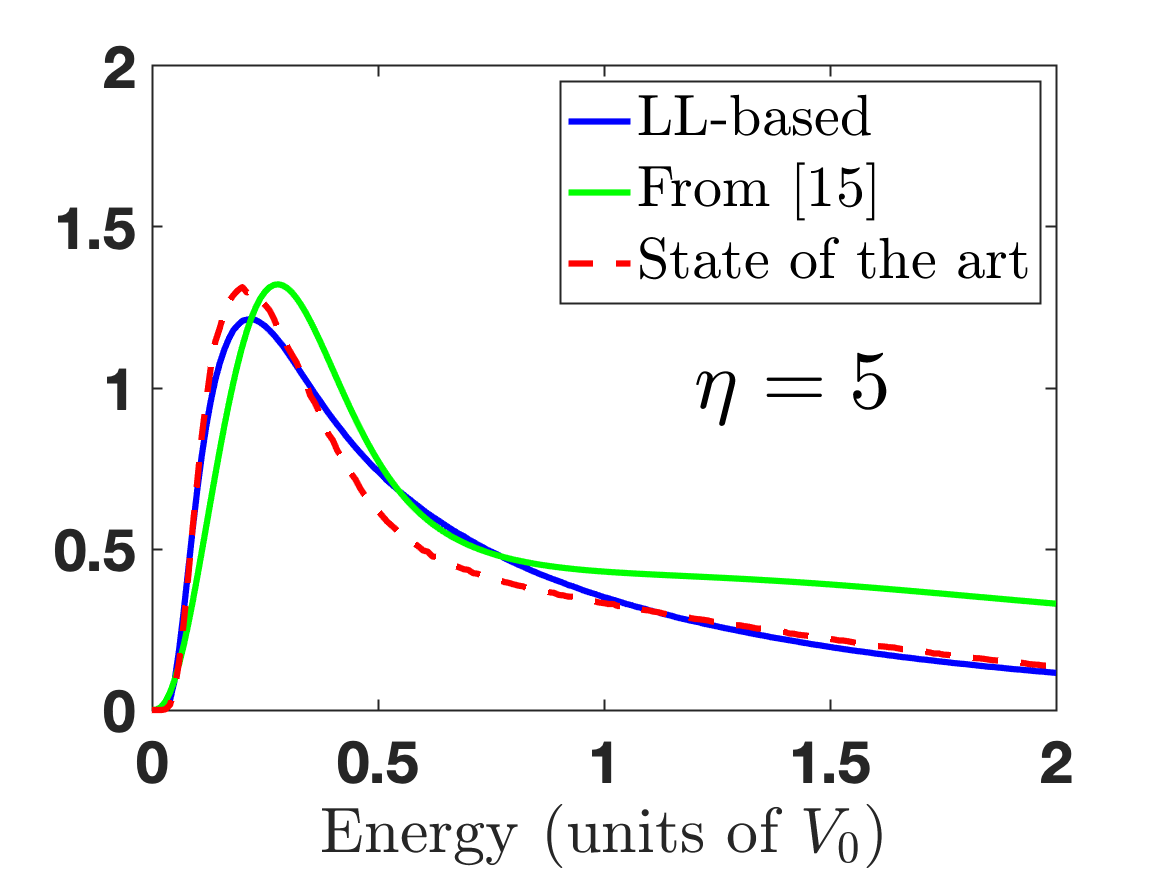}\\
\includegraphics[width=.49\columnwidth]{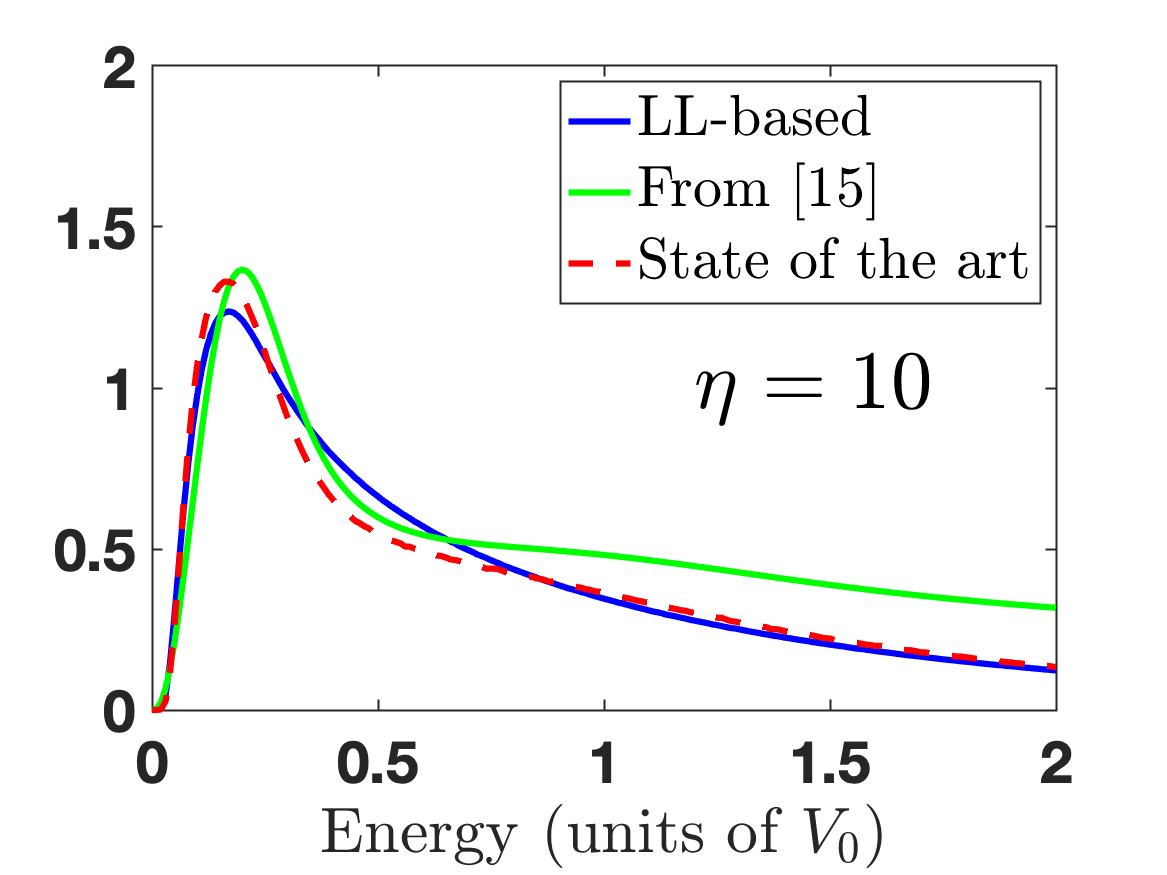}
\includegraphics[width=.49\columnwidth]{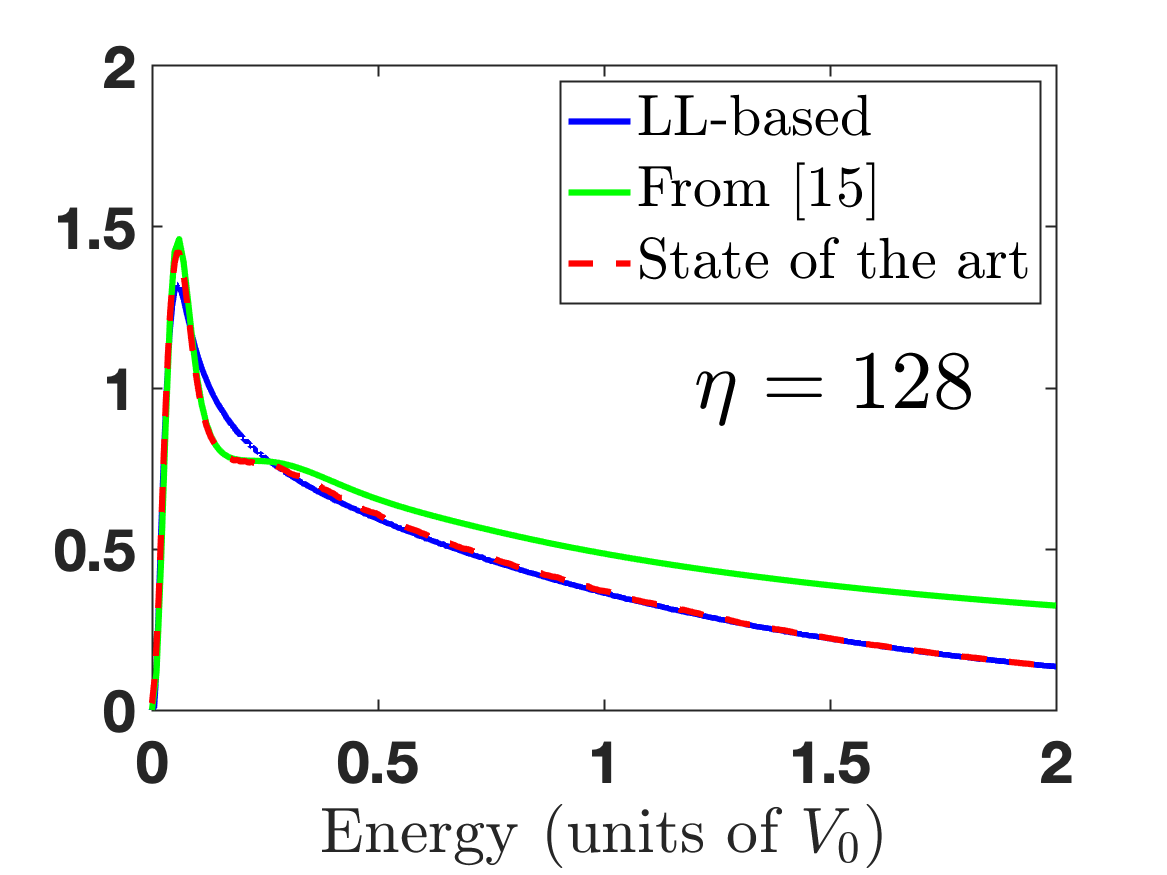}
\caption{Comparison between the spectral functions computed using 3 different methods: an exact method based on solving the Schrödinger equation (red dashed line), the LL-based approach (blue line) and the semiclassical expansion from~\cite{Prat2016} (green line). The comparison is performed for 6 values of $\eta$: 0.1, 0.5, 1, 5, 10, and 128. All three curves are very close in the semiclassical limit, but the LL-based estimate is more accurate in the quantum regime.}
\label{fig:1DPrat}
\end{figure}

To test the interest of the LL-based approach, we compare the results displayed in Fig.~\ref{fig:1DspeckleGaussian} with a semiclassical expansion proposed in Eq.~(30) of~\cite{Prat2016}, in which the wells of the original potential are approximated as a set of independent harmonic wells of random depths and curvatures. Figure~\ref{fig:1DPrat} displays the comparison of the spectral functions estimated using the exact computation (red dashed line), the LL-based approach (blue line), or the semiclassical approximation (green line), for 5 different values of $\eta$: 0.1, 0.5, 1, 5, 10, and 128. Among the first five values, four correspond to values used in Fig.~\ref{fig:1DspeckleGaussian}. The sixth value is the value shown in \citep{Prat2016}. We observe that this approximation~\cite{Prat2016} is very good for $\eta=128$ (and indeed more accurate at low energy than the LL-based estimate). However, for all other values ($\eta=10$ and below), the LL-based estimate is clearly closer to the actual function on a larger range of energies.

Finally, we explore potentials with a different distribution of values, namely a Gaussian distribution with also Gaussian spatial correlation, whose characteristics are:
\begin{align}\label{eq:Gaussianpotential}
\begin{cases}
    P(V) &= \displaystyle \frac{1}{\sqrt{2\pi} \, V_0} \, \exp(-\frac{V^2}{2V_0^2}) \\
    \overline{V} &= 0 \\
    g(x) &= \displaystyle V_0^2 \, \exp(-\frac{x^2}{2\sigma^2}) \,.
\end{cases}
\end{align}

\begin{figure}[ht!]
\centering
\includegraphics[width=.49\columnwidth]{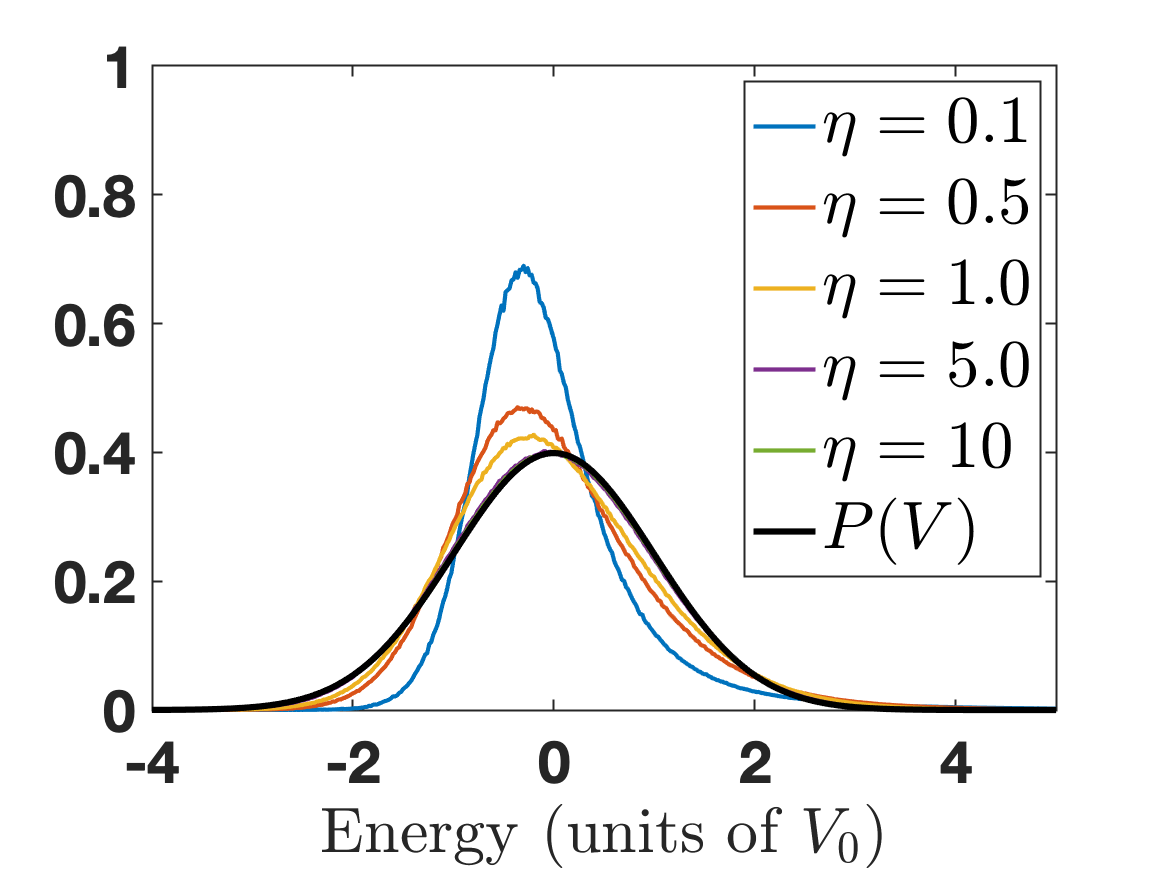}
\includegraphics[width=.49\columnwidth]{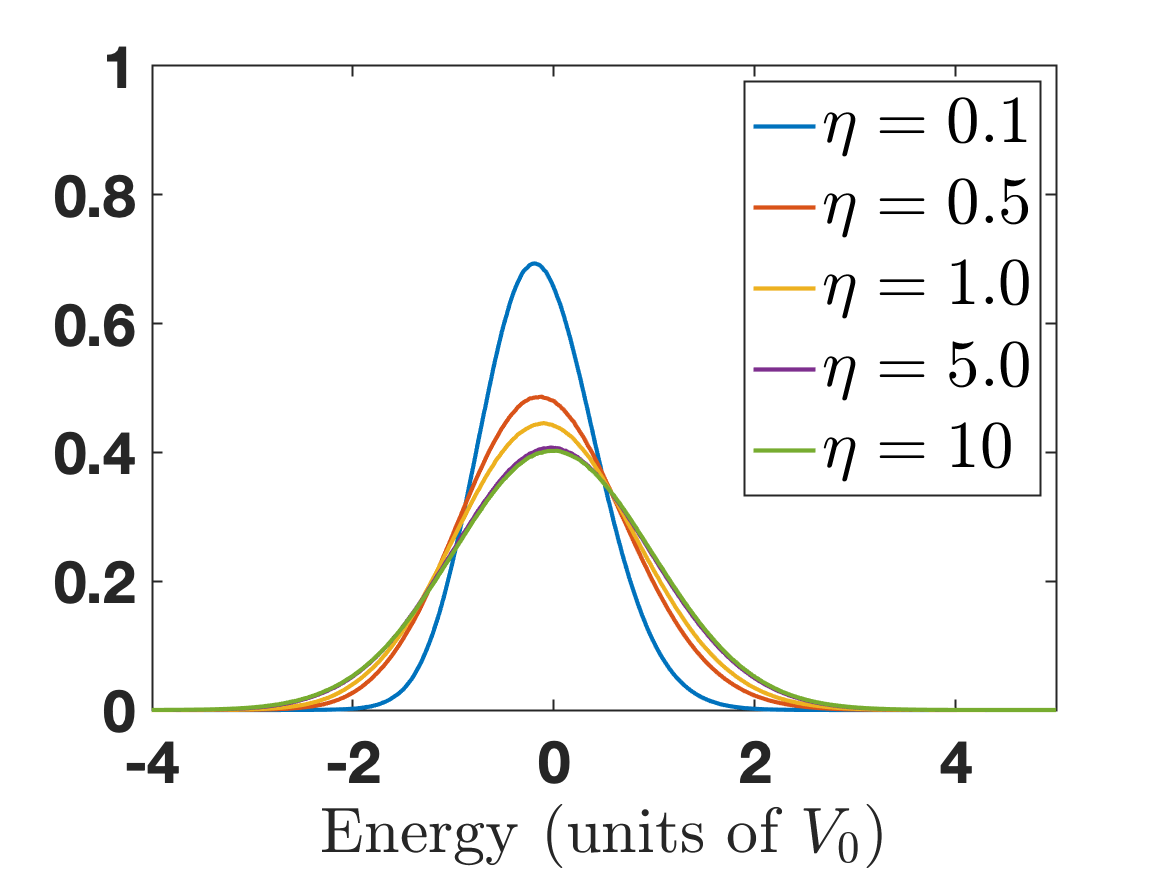}\\
\includegraphics[width=.49\columnwidth]{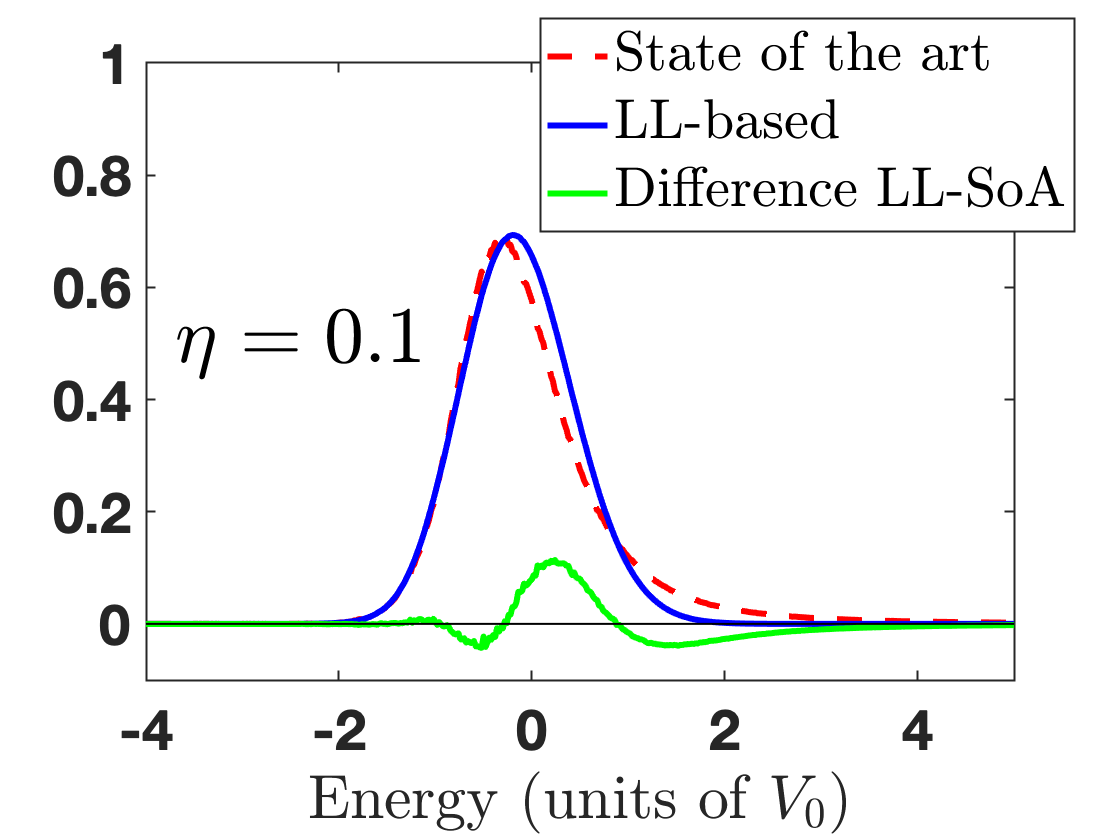}
\includegraphics[width=.49\columnwidth]{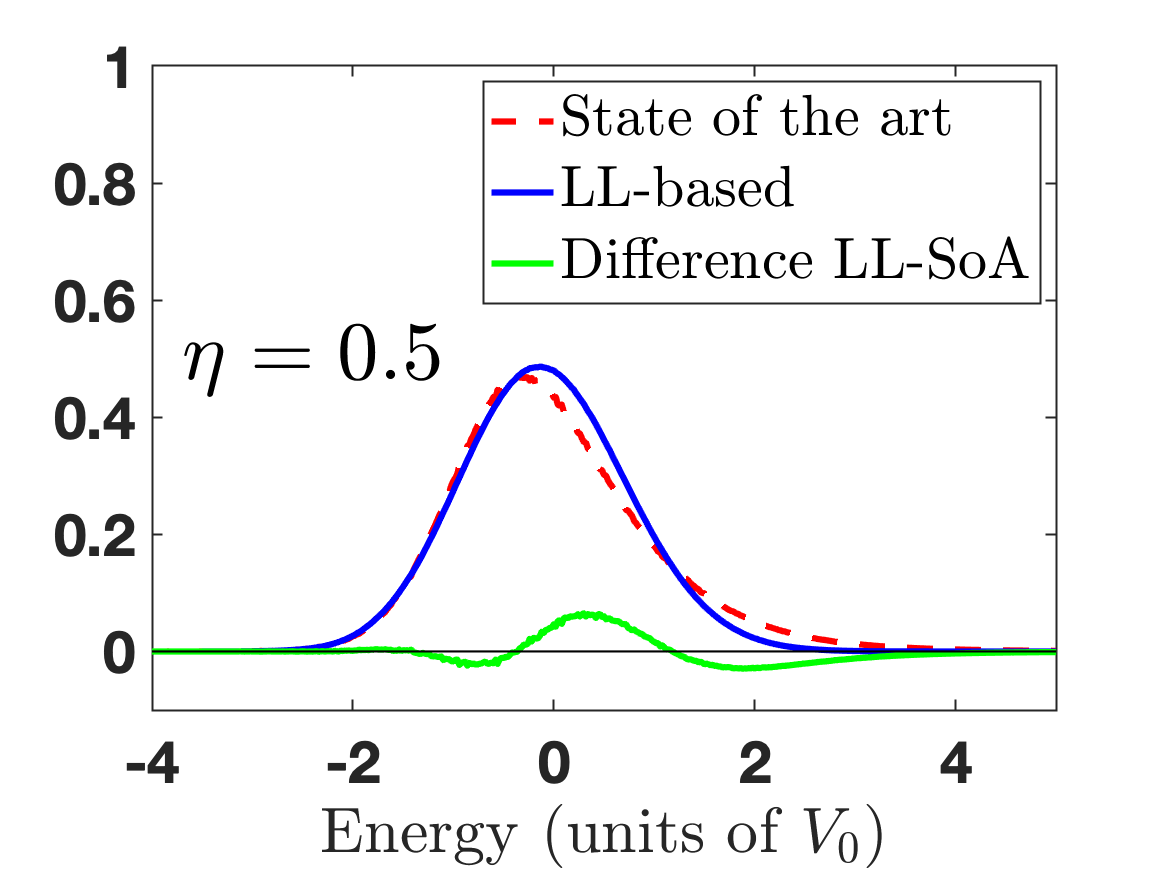}\\
\includegraphics[width=.49\columnwidth]{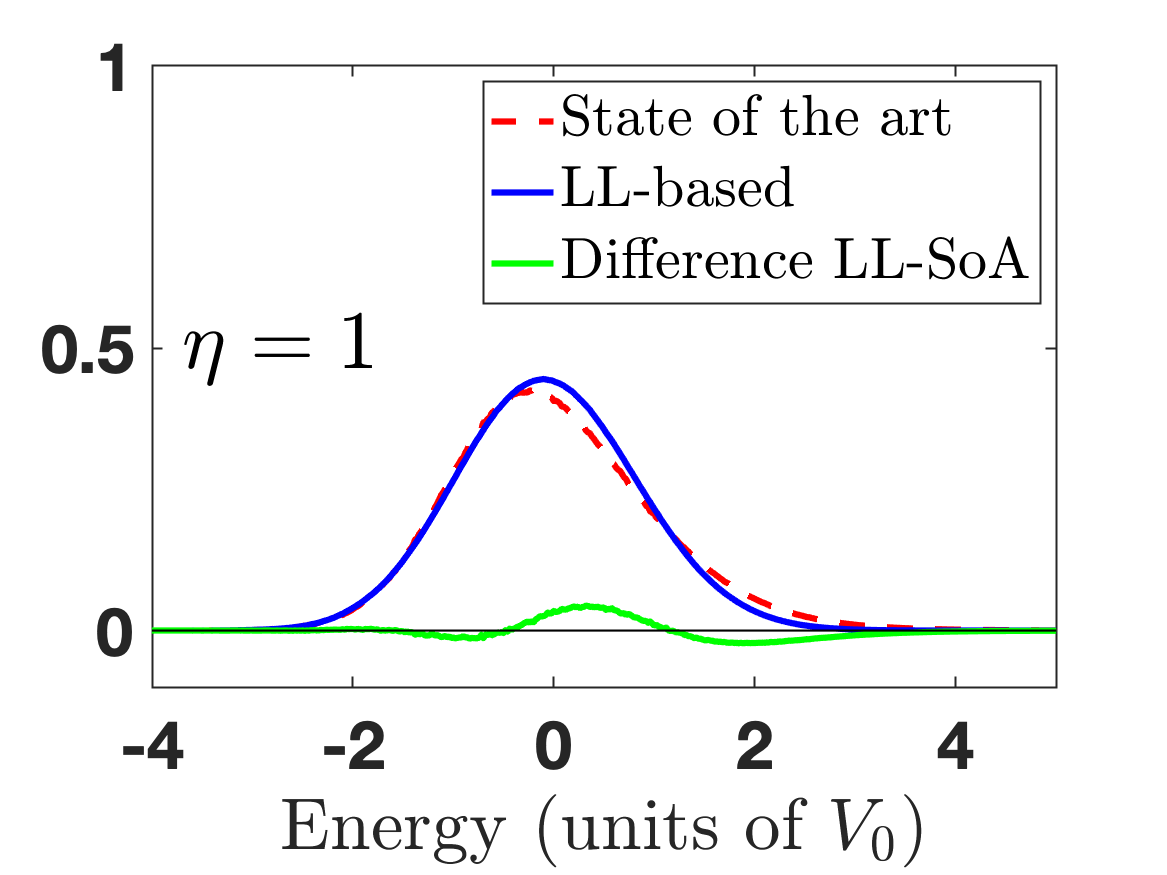}
\includegraphics[width=.49\columnwidth]{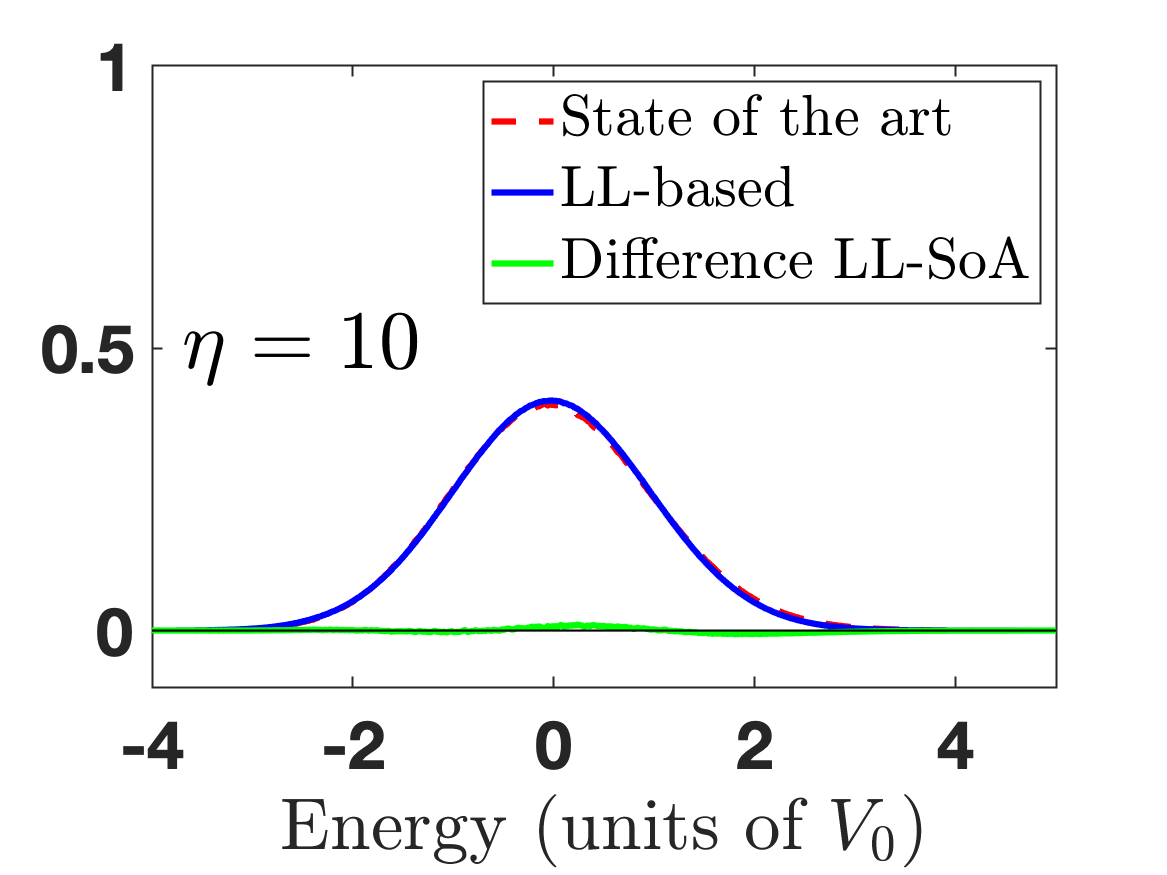}
\caption{Top left: Spectral functions (in units of $1/V_0$) in the case of a disordered potential with Gaussian-distributed values and Gaussian correlations [see Eq.~\eqref{eq:Gaussianpotential}], computed from the solutions to the Schrödinger equation for 5~different values of the parameter $\eta$ ranging from 0.1  to 10 (0.1, 0.5, 1, 5, and 10). The distribution of the values of the original potential~$V$ is also plotted in black. Top right: Spectral functions computed from the LL theory for the same values of $\eta$. Middle and bottom panels: Direct comparisons between the spectral functions computed using the Schrödinger equation (dashed red) and the LL theory (blue), for $\eta=0.1$, $0.5$, $1$, and $10$, respectively. The difference between both curves (the error) is plotted in green.}
\label{fig:1DGaussian}
\end{figure}

In order to properly define the LL in Eq.~\eqref{eq:landscape}, the eigenvalues of the Hamiltonian must all be positive. While this condition was naturally satisfied in the previous cases since the potential was always positive-valued, a problem arises for this new potential. In order to get around this difficulty, the potential is shifted for each realization by a quantity $-E_0 +\varepsilon$, where $E_0$ is the ground-state energy of the original potential and $\varepsilon$ is a small numerical value, chosen here to be equal to $0.1\,V_0$. After solving the landscape equation~\eqref{eq:landscape}, the values of the effective potential are then shifted back by the same amount. The dependence of our results on $\varepsilon$ was found to be negligible. The spectral functions and the LL-based estimates are averaged over 50,000 realizations (see Fig.~\ref{fig:1DGaussian}). We find here also a very good agreement between the LL-based estimates and the exact spectral functions.

However, we observe that the spectral functions have a very different shape. This is expected as, in the semiclassical limit, the spectral function at $k=0$ is simply the distribution of the values of the original potential~$V$, \eqref{eq:Gaussianpotential}. In the deep quantum regime, $\eta\ll 1$, the spectral function is qualitatively similar, a close-to-symmetric distribution looking more or less like a Gaussian function. Here again, the LL-based estimate reproduces the spectral function in all regimes (semiclassical, intermediate, quantum) without any adjustable parameter. By intrinsically accounting for the confinement energy inside the wells of the original potential, and the tunneling effect across its barriers, the LL perfoms a renormalization of the disorder. The difference between the exact computation and the LL estimate consists mostly of a very slight mismatch between the peaks, and decays very rapidly as one approaches the semiclassical limit $\eta \rightarrow +\infty$.
It is interesting in this situation to compare the LL-based results to the semiclassical approach proposed in~\cite{Trappe2015}, in which the spectral functions are computed by assessing the leading quantum corrections to the deep classical limit by the Wigner–Weyl formalism. The prediction is~\footnote{This formula results from Eqs.~(18), (50) and (55) in Ref.~\cite{Trappe2015}, with a small change for the correction of Eq.~(55) which is twice smaller in our one-dimensional system, compared to the two-dimensional case studied in~\cite{Trappe2015}}:
\begin{equation}
A_{\vb{0}}(E) = \frac{1}{\sqrt{2\pi} \, V_0} \, \exp(-\frac{V^2}{2V_0^2}) \left(1 - \frac{E(3V_0^2-E^2)}{12V_0^4}\right) \,.
\end{equation}

\begin{figure}[H]
\centering
\includegraphics[width=.49\columnwidth]{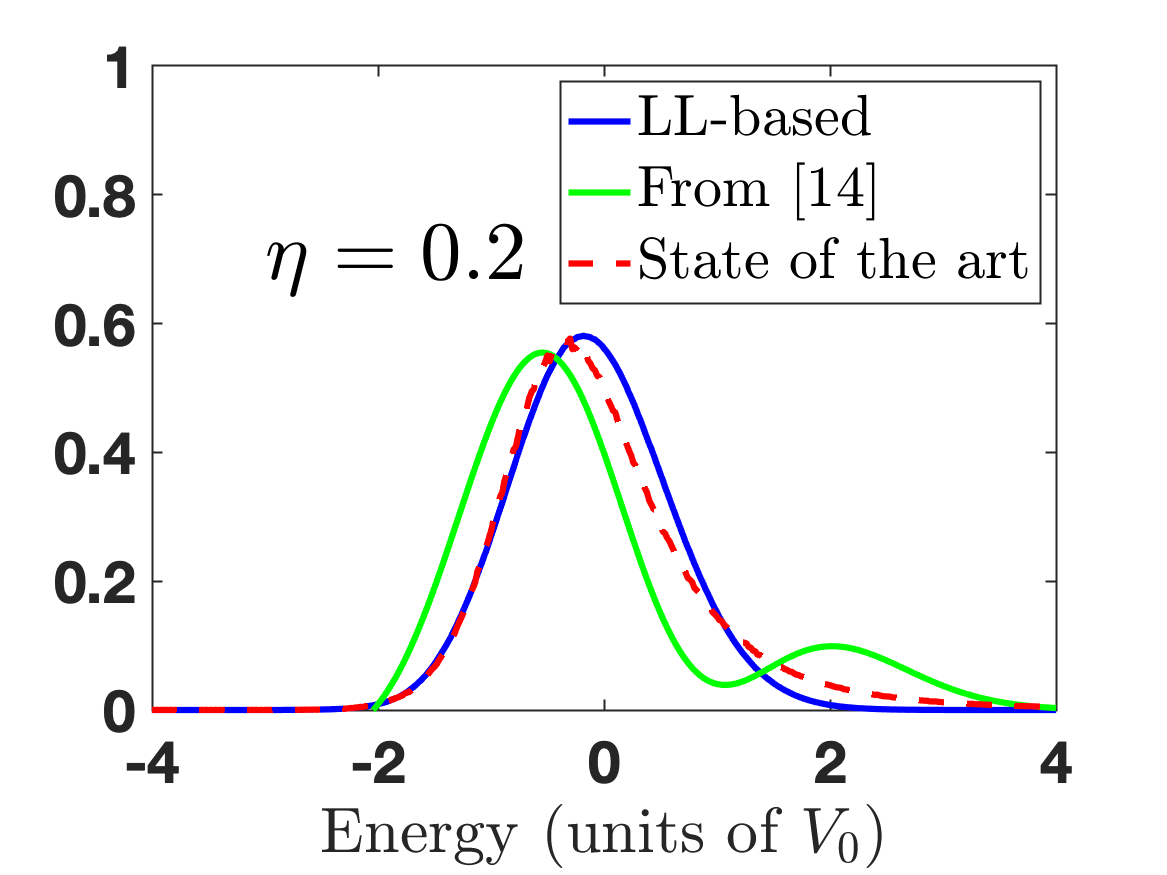}
\includegraphics[width=.49\columnwidth]{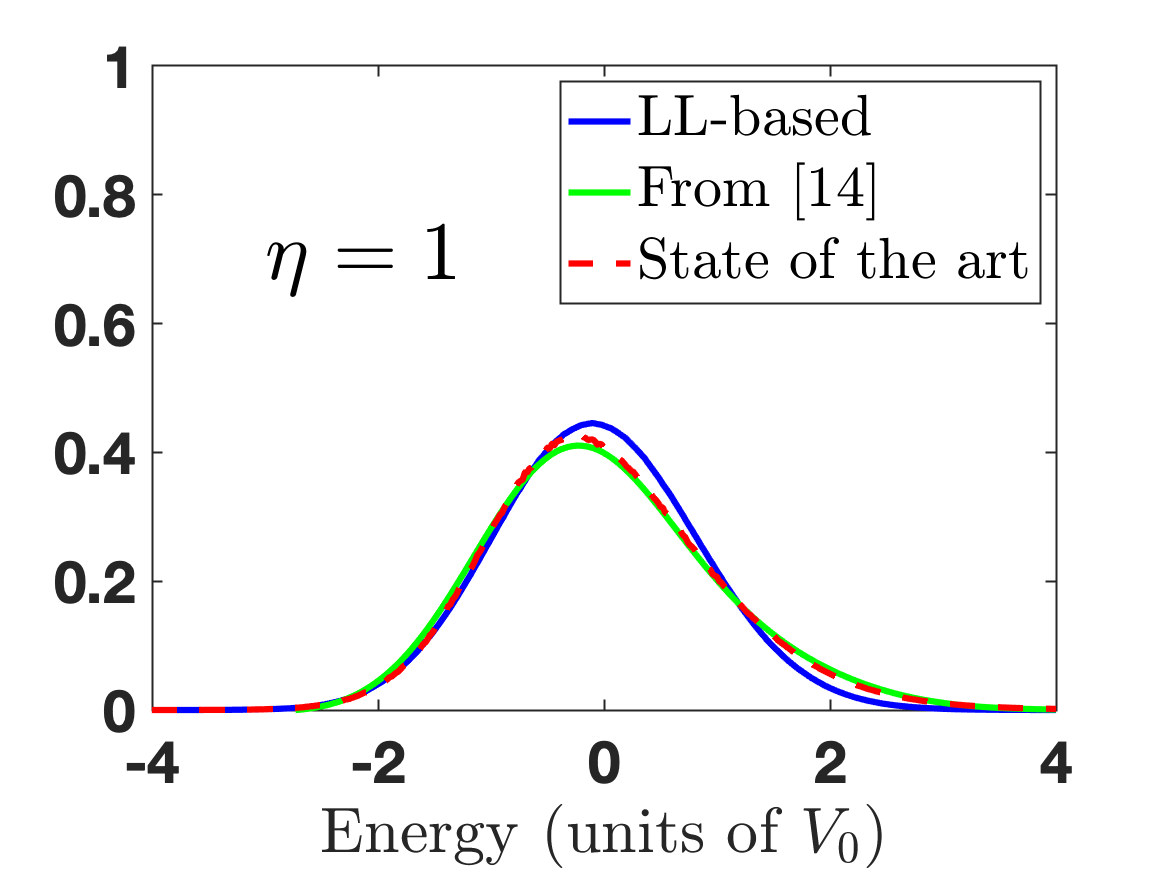}\\
\includegraphics[width=.49\columnwidth]{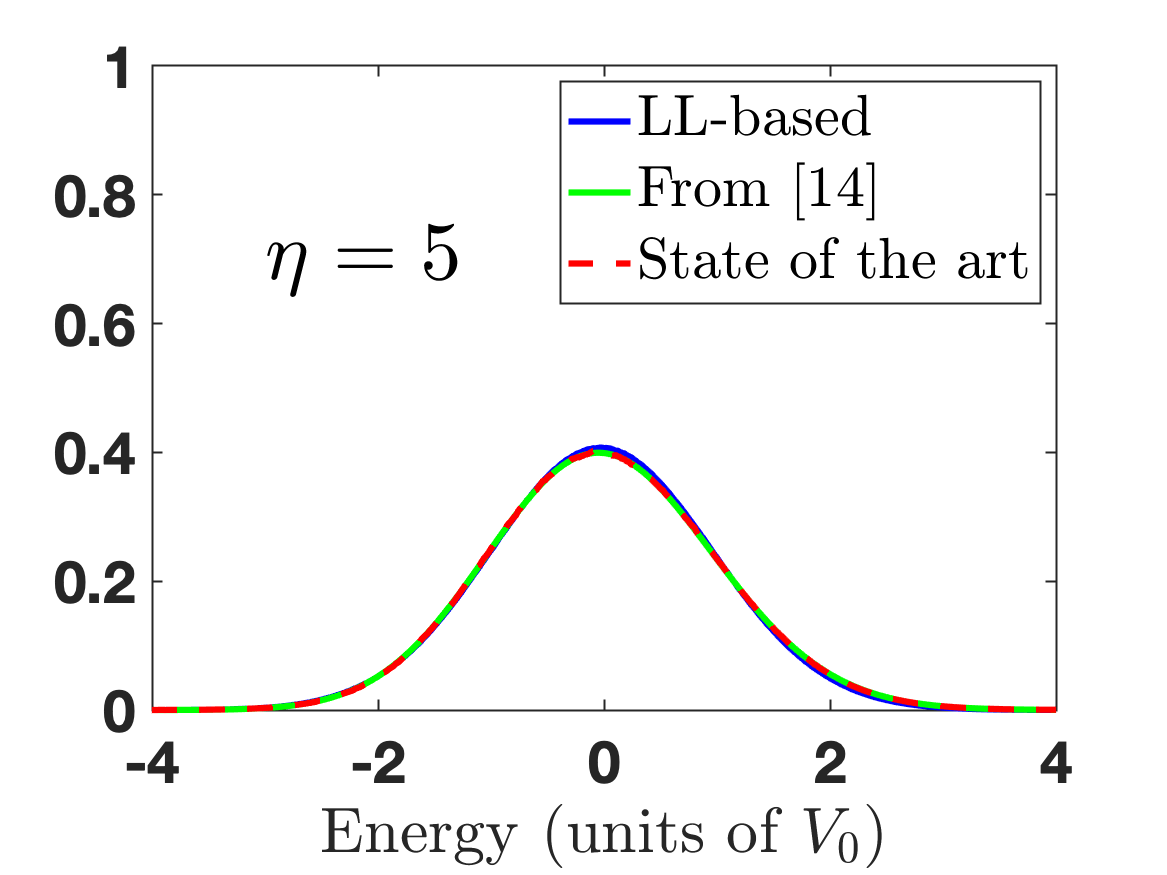}
\includegraphics[width=.49\columnwidth]{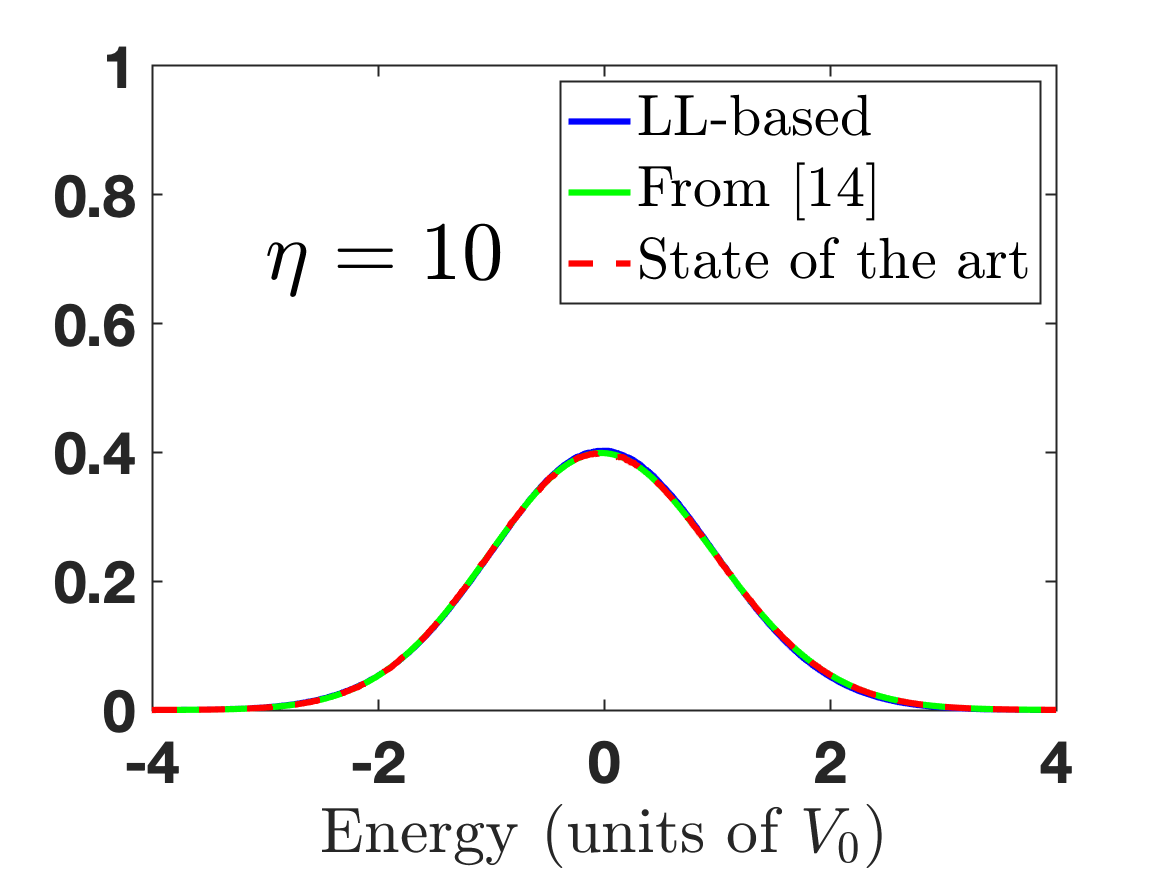}
\caption{Comparison between the spectral functions computed using 3 different methods: an exact method based on solving the Schrödinger equation (red dashed line), the LL-based approach (blue line) and the semiclassical expansion from~\cite{Trappe2015} (green line). The comparison is performed for 4 values of $\eta$: 0.2, 1, 5, and 10. All three curves are almost identical in the semiclassical limit, but the LL-based approach remains closer to the actual spectral function in the quantum regime.}
\label{fig:1DTrappe}
\end{figure}

The comparison is carried out in Fig.~\ref{fig:1DTrappe}. We observe here that both the LL-based and the semiclassical computations are very close to the actual spectral functions in the semiclassical regime. However, the LL-based computation is closer to the actual spectral function at low value of $\eta$, i.e., in the quantum regime, while the semiclassical approximation appears slightly closer to the true spectral function in the intermediate regime ($\eta \approx 1$).

\section{Conclusion}

{In this paper, we build on the previous results in the localization landscape theory~\cite{Arnold2016}, which connected the density of states with the classical Hamiltonian $H_1(\vb{x},\vb{k}) = \hbar^2\vb{k}^2/2m + V_u(\vb{x})$. We propose a new version of the Wigner-Weyl approach which uses the level sets of $H_1(\vb{x},\vb{k})$ rather than those of the classical Hamiltonian~$H(\vb{x},\vb{k})$. The resulting approximation of the spectral function covers all regimes, from the deep quantum regime where the disorder is small to the semiclassical regime where it dominates the dynamics, through the intermediate regime. This makes it possible to compute efficiently the spectral function---here tested for momentum $\vb{k}=\vb{0}$---for various types of disorder.

Although we presented here results for one-dimensional systems, the method is expected to work also in higher dimensions. Moreover, the similarity of the differences between the exact computations and the LL-based formula for every value of $\eta$ opens the way to look for a universal correction.

As the knowledge of the spectral function is a key ingredient for studying transport properties and, for example, the onset of localization, it may well be that the present work will lead to rather simple calculations of e.g. the mobility edge in three-dimensional systems, a notoriously difficult problem for spatially correlated potentials~\cite{Piraud2013,Yedjour2010,Delande2014,Pasek2017}.

% The \nocite command causes all entries in a bibliography to be printed out
% whether or not they are actually referenced in the text. This is appropriate
% for the sample file to show the different styles of references, but authors
% most likely will not want to use it.
%\nocite{*}

%\bibliographystyle{apsrev4-2}
\bibliography{Pelletier_spectral}

\end{document}